\UseRawInputEncoding 
\documentclass[aps,pre,twocolumn,superscriptaddress,showpacs,floatfix,longbibliography]{revtex4-1}

\makeatletter
\def\@bibdataout@aps{%
\immediate\write\@bibdataout{%
@CONTROL{%
apsrev41Control%
\longbibliography@sw{%
    ,author="08",editor="1",pages="1",title="0",year="1"%
    }{%
    ,author="08",editor="1",pages="1",title="",year="1"%
    }%
  }%
}%
\if@filesw \immediate \write \@auxout {\string \citation {apsrev41Control}}\fi
}
\makeatother

\usepackage{graphicx}% Include figure files
\usepackage{dcolumn}% Align table columns on decimal point
\usepackage{bm}% bold math
\usepackage{multirow}

\usepackage{amssymb}
\usepackage{amsmath}
\usepackage{graphicx}
\usepackage{xcolor}
\usepackage[colorlinks,citecolor=blue,linkcolor=red,urlcolor=blue]{hyperref}
\newcommand{\revone}[1]{{ \color{black} #1}}
\newcommand{\revtwo}[1]{{ \color{black} #1}}
\newcommand{\yurong}[1]{{ \color{black} #1}}

\begin{document}
\title{Dual dynamic scaling in deconfined quantum criticality}
\author{Yu-Rong Shu}
\affiliation{School of Physics and Materials Science, Guangzhou University, Guangzhou 510006, China}

\author{Shuai Yin}
\email{yinsh6@mail.sysu.edu.cn}
\affiliation{School of Physics, Sun Yat-Sen University, Guangzhou 510275, China}
\date{\today}

\begin{abstract}
Emergent symmetry is one of the characteristic phenomena in deconfined quantum critical point (DQCP). As its nonequilibrium generalization, the dual dynamic scaling was recently discovered in the nonequilibrium imaginary-time relaxation dynamics in the DQCP of the $J$-$Q_3$ model. In this work, we study the nonequilibrium imaginary-time relaxation dynamics in the $J$-$Q_2$ model, which also hosts a DQCP belonging to the same equilibrium universality class. We not only verify the universality of the dual dynamic scaling at the critical point, but also investigate the breakdown and the vestige of the dual dynamic scaling when the tuning parameter is away from the critical point. We also discuss its possible experimental realizations in devices of quantum computers.
\end{abstract}

\maketitle

%%%%%%%%%%%%%%%%%%%%%%%%%%%%%%%%%%%%%%%%%%%%%%%%%%%%
\section{Introduction}
Symmetry plays pivotal roles in identifying, characterizing and classifying phases and phase transitions in condensed matter physics~\cite{Landaubook,Sachdevbook}. Traditional Landau-Ginzburg-Wilson (LGW) theory of phase transitions is based on the mechanism of spontaneous symmetry breaking, in which the symmetry of the ordered phase is always lower than that of the microscopic model~\cite{Landaubook}. In contrast, many critical systems show higher symmetry in the infrared limit than they do in the ultraviolet limit. For instance, spacetime supersymmetry can emerge at the critical point in some topological materials~\cite{Leess2007prb,Sheng2014science,Yao2015prl,Berg2015prl,Affleck2015prl,Maciejko2016prl,Yao2017prl,Yao2017prl2,Fendley2018prl,Liucx2019prb,Yin2020prb,Moon2021prb}; Lorentz symmetry can emerge in the superfluid-Mott insulator quantum phase transition~\cite{Pollet2012prl,Pollet2013prl}, critical two-subband quantum wires~\cite{Garst2009prl}, and phase transitions in Dirac systems~\cite{Yao2017nc,Herbut2016jhep,Michael2018prb1}; $\mathbb{SU}(3)$ symmetry can emerge in the critical spin-$2$ chain with translational invariant interaction and in the critical spin-$1$ chain with random bond interaction~\cite{PCChen2014prl,Quito2015prl}; extended $\mathbb{O}(N)$ symmetry can emerge at the multicritical point~\cite{Vicari2003prb,Scherer2018prb,Roy2018prb,Scherer2020prr}. In two-dimensional (2D) spin systems, a prominent example in which the emergent symmetry arises as its characteristic phenomenon is the deconfined quantum critical point (DQCP)~\cite{Senthil2004science,Sachdev2004prb,Senthil2004prb}. The DQCP was proposed as a new mechanism of continuous phase transition between two spontaneous symmetry breaking phases, while the usual LGW paradigm asserts that this kind of phase transition should be first ordered~\cite{Senthil2004science,Sachdev2004prb,Senthil2004prb,Hux2005prl,Kleinert2005prl,Matthew2006prb,Sandvik2007prl,Sudbo2007prb,Melko2008prb,Kaul2008prl,Mila2008prb,Kleinert2008prb,Sandvik2009prb,Sandvik2009prb1,Sandvik2010prl,Sandvik2011prl,Nahum2011prl,Bartosch2013prb,Sudbo2013epl,Sandvik2013prl,Damle2015prb,Nahum2015prx,Sandvik2016science,Kaul2017prl,Meng2017prx,Wang2017prx,Sandvik2017prx,Heyc2017prb,Youyz2018prx,Sachdev2018prx,Zhangxf2018prl,Xuck2018prb,GazitE6987,Guowa2019nc,Li2019,Michael2019prb,Janssen2020prb,Sandvik2020iop,Zayed2017naturephysics}. It was shown that for the DQCP separating the \revone{antiferromagnetic (AFM) phase and valence-bond-solid (VBS) phase} in $\mathbb{SU}(2)$-invariant quantum magnets, $\mathbb{SO}(5)$ symmetry emerges to reconcile VBS and AFM order parameters~\cite{Sandvik2007prl,Sandvik2009prb1,Nahum2015prl,Nahum2019prl}; while for the DQCP in easy-plane quantum magnets, $\mathbb{SO}(4)$ symmetry emerges~\cite{Mans2019prl}.

%two phases with non-compatible broken symmetries
The emergent symmetry and its breaking give rise to the intriguing critical properties in DQCP. For the square lattice, on the VBS side, the emergent continuous symmetry breaks down to discrete $\mathbb{Z}_4$ symmetry~\cite{Senthil2004science,Sachdev2004prb,Senthil2004prb,Sandvik2009prb1}. Accordingly, the fugacity of the quadrupled coherent monopoles works as a dangerously irrelevant scaling variable, which is irrelevant exactly at the critical point, but relevant in the ordered VBS phase~\cite{Oshikawa2000prb,Louj2007prl,Kawashima2015prb,Shao2020prl,Shao2021prb,Delamotte2015prl}. Pertinent to this variable, an extra divergent length $\xi'$, which measures the spinon confinement length or the thickness of the VBS domain walls, develops, in addition to the conventional correlation length $\xi$~\cite{Senthil2004prb,Sandvik2016science}. And they satisfy $\xi'\propto \xi^{\nu'/\nu}$ with $\nu$ and $\nu'$ being the corresponding critical exponents~\cite{Oshikawa2000prb,Louj2007prl,Kawashima2015prb,Shao2020prl,Shao2021prb,Delamotte2015prl}. It was plausibly shown that the interplay between these two length scales may take responsibility for some anomalous equilibrium scaling behaviors near the DQCP~\cite{Nahum2015prx,Sandvik2016science}, although a very weak first-order phase transition with pseudo-critical phenomena cannot be ruled out~\cite{Kuklov2005ptps,KUKLOV20061602,Kuklov2008prl,Jiang2008iop,Chenk2013prl,Zan2018jhep,Zan2018scipost,Wangc2020prb,Nahum2020prb,Lauchli2021}.

On the other hand, from the inflating universe to the flowing rivers, equilibrium phenomena are just the exception rather than the rule in nature. Moreover, investigations on nonequilibrium critical properties are of particular significance since universal time-dependent behaviors always appear near a critical point~\cite{Hohenberg1977rmp,Dziarmaga2010review,Polkovnikov2011rmp,Rigol2016review,Mitra2018arcmp}. In classical systems, the theory of critical dynamics has been well established by classifications of the dynamic universality classes~\cite{Hohenberg1977rmp}. Recently, spurred by the remarkable experimental progresses in manipulating and detecting the nonequilibrium quantum process, the quantum critical dynamics has attracted intensive attention from both theoretical and experimental aspects~\cite{Dziarmaga2010review,Polkovnikov2011rmp,Rigol2016review,Mitra2018arcmp}. Among these studies, it was shown that the scaling properties of the imaginary-time relaxation dynamics near quantum critical points resemble those in the classical short-time critical dynamics~\cite{Janssen1989,Lizb1995prl,Zhengb1996prl,Albano2011iop,Yins2014prb,Yins2014pre,Shu2017prb,Shu2020prb,Zhong2021prb}.

Inspired by above intriguing issues, a nature question arises: how the emergent symmetry and the associated scaling form with two length scales affect the nonequilibrium dynamics in DQCP. In our previous work~\cite{Shu2021prl}, we studied the imaginary-time relaxation dynamics at the DQCP of the $J$-$Q_3$ model. We found that with an ordered initial VBS (N\'{e}el) state, the relaxation dynamics of the VBS (N\'{e}el) order parameter is controlled by the conventional correlation length $\xi$, while the dynamics of the N\'{e}el (VBS) order parameter is controlled by the confinement length $\xi'$. A \textit{dual dynamic scaling}, which states that the dynamic scaling forms change to their dual partners as the initial states are changed to their dual counterpart, is then proposed. This dual dynamics scaling can be regarded as the nonequilibrium incarnation of the emergent symmetry in the equilibrium case. Given the fact that emergent symmetry is a common phenomenon in DQCP~\cite{Senthil2004science,Sachdev2004prb,Senthil2004prb,Sandvik2009prb1,Nahum2015prl,Nahum2019prl,Mans2019prl}, it is imperative to explore the universality and robustness of the dual dynamic scaling in other models that host a DQCP.

In this paper, we study the nonequilibrium imaginary-time relaxation dynamics in the $J$-$Q_2$ model~\cite{Sandvik2007prl}.
\revtwo{The model shares the same equilibrium universality class with the $J$-$Q_3$ model but has a weaker VBS order~\cite{Sandvik2009prb,Tang2011prl}.}
After estimating the critical point via the dynamics of the sign function of the order parameters, we study the relaxation behaviors of the N\'{e}el and VBS order parameters from different initial states at the critical point. By comparing the scaling forms for different quantities, we verify the universality of the dual dynamic scaling by showing that (i) the N\'{e}el and VBS order parameters are controlled by different length scales for different initial states; (ii) the dynamic scaling forms exchange under the \revtwo{exchange} of the initial states, similar to the case of the $J$-$Q_3$ model. Moreover, we investigate dynamic scaling properties when the tuning parameter is away from the critical point. Strikingly, we find that in the short-time stage, the dual dynamic scaling, with a proper generalization of the dual transformation, can exist even in the presence of the off-critical-point effects, although in equilibrium the emergent symmetry fades away \revtwo{once the system is set away from its critical point}. A possible experimental realization based on programmable quantum devices is also discussed.

The rest of the paper is organized as follows. In Sec.~\ref{modeldy}, we introduce the equilibrium properties of the $J$-$Q_2$ model, the protocol of the imaginary-time relaxation dynamics and the numerical method used, Then, in Sec.~\ref{reviewscaling}, we give a brief review on the nonequilibrium dynamic scaling form. After estimating the critical point of the $J$-$Q_2$ model via the nonequilibrium scaling in Sec.~\ref{criticalpoint}, we explore the dual dynamic scaling at the critical point for various initial states in Sec.~\ref{dualscaling}. In Sec.~\ref{offdualscaling}, we show the dynamic scaling behavior with the off-critical-point effects. Then we discuss the experimental realizations in Sec.~\ref{experimentdis}. A summary is given in Sec.~\ref{summaryd}.

\section{\label{modeldy}Model and Imaginary-time Relaxation Dynamics}
The Hamiltonian of the Sandvik's $J$-$Q_2$ model reads~\cite{Sandvik2007prl}
\begin{equation}
\label{eq:hamiltonian}
H=-J\sum_{\langle ij\rangle}P_{ij}-Q\sum_{\langle ijkl\rangle}{P_{ij}P_{kl}},
\end{equation}
in which $J>0$ and $Q>0$, $\langle ij\rangle$ and $\langle ijkl\rangle$ denote nearest neighbors and two nearest-neighbor pairs in horizontal \revtwo{rows} or vertical columns on the square lattice, respectively, and $P_{ij}$ denotes the spin singlet operator defined as $P_{ij}\equiv\frac{1}{4}-{\bf S}_{i}\cdot{\bf S}_{j}$ with $\bf S$ being the spin-$1/2$ operator. The system favors the N\'{e}el phase with a finite order parameter ${\bf M}\equiv \sum_r{(-1)^{r}\bf S}_r/N$ when \yurong{$q\equiv J/Q\gg q_{\rm c}$}, while it favors the VBS phase with a finite $\bf D$ when \yurong{$q\ll q_{\rm c}$}~\cite{Sandvik2007prl}, in which ${\bf D}=D_x\hat{x}+D_y\hat{y}$ with $D_{x(y)}\equiv \sum_{r}(-1)^{r_{x(y)}}{\bf S}_r\cdot {\bf S}_{r+\hat{x}(\hat{y})}/N$. $N=L^{d}$ is the number of spins and $\hat{x}$, $\hat{y}$ denote the unit lattice vector in the $x$ and $y$ direction, respectively.
These two ordered phases break different symmetries: the N\'{e}el order breaks the spin rotation symmetry, while the VBS order breaks the translation symmetry. The phase transition between them happens at $q=q_{\rm c}\simeq 0.045$~\cite{Sandvik2007prl,Sandvik2016science,Sandvik2020iop}. According to the LGW theory, this phase transition should be first ordered~\cite{Landaubook}. However, plenty of numerical results with scrutiny demonstrate that this phase transition is a continuous one satisfying the DQCP theory~\cite{Sandvik2007prl,Sandvik2010prl,Sandvik2016science,Sandvik2020iop}.
Here we list the critical exponents of the $J$-$Q_2$ model relevant to this study. The dynamic exponent $z$ is equal to $1$~\cite{Sandvik2007prl}.
Recent works show that the anomalous dimension $\eta\simeq 0.25$~\cite{Nahum2015prl}, leading to $(2\beta/\nu)\simeq 1.25$ from the scaling laws and the correlation length exponent is $\nu\simeq 0.455$~\cite{Sandvik2020iop}.

\begin{figure}[tbp]
\centering
  \includegraphics[width=\linewidth,clip]{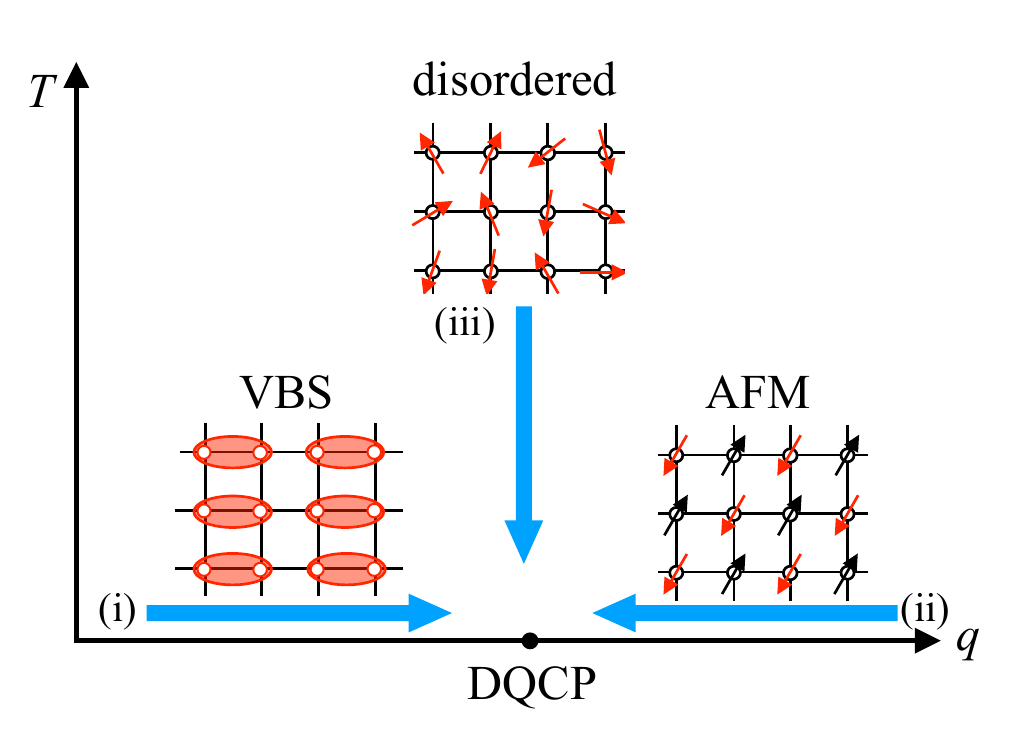}
  \vskip-3mm
  \caption{Sketch of the quench dynamics in imaginary-time with different initial states. The initial states are prepared as (i) the saturated VBS state, (ii) the saturated AFM state, and (iii) the completely disordered initial state. All these states have vanishing correlation length and correspond to the fixed points of the initial states under the renormalization group transformation.
  }
  \label{fig:quench}
\end{figure}

Remarkably, it was shown that an emergent $\mathbb{SO}(5)$ symmetry appears at the critical point of this N\'{e}el-VBS transition~\cite{Matthew2006prb,Nahum2015prl,Nahum2019prl}. This symmetry is induced by the conservation of the monopole defects~\cite{Motrunich2004prb}, indicating that the critical point is described by the non-compact $(2+1)$D quantum electrodynamics with the deconfined spinons as its matter field~\cite{Senthil2004science,Sachdev2004prb,Senthil2004prb}. Moreover, in the VBS phase, confined spinon-pairs are formed and the discrete $\mathbb{Z}_4$ symmetry is broken~\cite{Haldane1988prl,Read1990prb,Sachdev1991prl}. To reconcile the emergent continuous symmetry at $q=q_{\rm c}$ and the discrete symmetry for $q<q_{\rm c}$, the fugacity of the monopole defects should take responsibility as a dangerously irrelevant scaling variable. Accordingly, besides the usual correlation length $\xi\propto |\delta|^{-\nu}$ \revone{with $\delta=q-q_{\rm c}$ the distance to the critical point}, the confinement length $\xi'\propto |\delta|^{-\nu'}$ with $\nu'\simeq 0.585$~\cite{Sandvik2016science}, measuring the averaged distance between two spinons, also plays significant roles~\cite{Oshikawa2000prb,Louj2007prl,Kawashima2015prb,Shao2020prl,Shao2021prb,Delamotte2015prl}. It was shown that the interplay between these two characteristic scales can explain the anomalous scaling behaviors for the domain wall density and the susceptibility~\cite{Sandvik2016science}.

For the imaginary-time relaxation dynamics, the evolution of the wave function $|\psi(\tau)\rangle$ obeys the imaginary-time Schr\"{o}dinger equation
\begin{equation}
\label{eq:dynamics}
-\frac{\partial}{\partial\tau}|\psi(\tau)\rangle=H|\psi(\tau)\rangle,
\end{equation}
with the normalization condition $Z=\langle\psi(\tau)|\psi(\tau)\rangle=1$~\cite{Yins2014prb,Yins2014pre}.
\revone{The formal solution of the Schr\"{o}dinger equation is given by
\begin{equation}
\label{eq:formal}
|\psi(\tau)\rangle = \frac{1}{Z}U(\tau)|\psi(\tau_0)\rangle,
\end{equation}
in which $U(\tau)\equiv \textrm{exp}(-\tau H)$ is the imaginary-time evolution operator and $\tau_0$ is the starting point of the evolution.}
\revone{
  In studies of relaxation dynamics, one focuses on the dynamical scaling behaviors of different quantities when relaxed from a given state. Therefore, the initial state is of key significance.
}
In the present work, as illustrated in Fig.~\ref{fig:quench}, we will consider three kinds of uncorrelated initial states: (i) the saturated VBS state, (ii) the saturated AFM state, and (iii) the completely disordered state. All these three states have vanishing correlation length and correspond to the fixed points of the initial states under the renormalization group transformation.
% The imaginary-time dynamics can be readily realized in the unbiased quantum MC simulation~\cite{Sandvik2010review,Shu2017prb,Shu2020prb}. The central idea of the projector QMC method is to perform series expansion of $U(\tau)\equiv \textrm{exp}(-\tau H)$ in the normalization. The expansion order can be truncated to some maximum length that causes no detectable error. The operator sequence and states are then importance-sampled and measurements can follow with the corresponding operator as input. We also implement the global loop-update scheme to improve the efficiency in the importance sampling procedure.

\revone{
  A wide range of quantum Monte Carlo (QMC) methods, including the stochastic series expansion, projector QMC, and world line methods, have natural connections to simulations of the imaginary-time evolution of quantum spin systems~\cite{Sandvik2010review}. In particular, the projector QMC has proven a powerful tool in pursuing the imaginary-time dynamics~\cite{Sandvik2010review,PolkovnikovSandvik2013prb,De_Grandi_2013,Farhi2012,Shu2017prb,Shu2020prb,Shao2015,Weinberg2017,Shu2021prl}.
  In the projector QMC method, the imaginary-time evolution operator is Taylor-expanded and the normalization can then be written as the sum of the operator sequence acting on some suitable basis states, such as the $S^{z}$ basis, the valence bond basis etc~\cite{Tang2011prl,Beach2006npb}. The sum over the operator sequence, along with the basis states, and the expansion power are then importance sampled. Local and global operator-loop updating schemes are developed to improve the efficiency of the Monte Carlo sampling. The expansion order is truncated to some maximum length, which is not strictly necessary in principle but brings significant convenience to implementations of the method. Note that such truncation of the expansion order causes no detectable errors. Expectation values of physical quantities are then estimated in the finial state $|\psi(\tau)\rangle$ that is propagated from $|\psi(\tau_0)\rangle$ with the sampled operator sequence and the corresponding basis states.
\revone{In QMC simulations, to achieve a given initial state for the system, one need to fix the boundaries of the imaginary-time propagation direction. Therefore, different basis are applied for convenience. For the saturated VBS state, the valence bond basis is used~\cite{Tang2011prl,Beach2006npb}, while for the AFM and disordered state, the $S^z$ basis is used~\cite{Farhi2012,Shu2017prb,Shu2020prb,Shu2021prl,Weinberg2017,PolkovnikovSandvik2013prb,De_Grandi_2013,Tang2011prl,Sandvik2010review}.
}
For more detailed introduction of the method, we refer to the literature~\cite{Sandvik2010review,PolkovnikovSandvik2013prb,De_Grandi_2013,Farhi2012,Shu2017prb,Tang2011prl,Sandvik2007prl,Sandvik2010prb}.
}

\section{\label{reviewscaling}Brief review of the dynamic scaling in $J$-$Q_3$ model}
To study the nonequilibrium imaginary-time critical dynamics, one should at first clarify the scaling relation between the imaginary-time $\tau$ and the correlation length $\xi$. For usual critical point with one single divergent length scale, the scaling relation between $\tau$ and $\xi$ satisfies $\xi\propto \tau^{1/z}$. In contrast, for the criticality with two length scales, there are two possibilities: (i) $\xi\propto \tau^{1/z}$ and $\xi'\propto \tau^{\nu'/\nu z}$; and (ii) $\xi'\propto \tau^{1/z}$ and $\xi\propto \tau^{1/z_u}$ with $z_u$ being $z_u\equiv z \nu'/\nu$ (The subscript $u$ means the usual correlation length). It has been shown that scenario (ii) is selected by the DQCP in the $J$-$Q_3$ model~\cite{Shu2021prl}. 

With scenario (ii), the imaginary-time dynamics for saturated ordered and completely disordered initial states should obey the scaling form
\begin{equation}
\label{eq:operator}
Y(\tau,\delta,L)=\tau^{\frac{s}{\tilde{z}}}f(\delta \tau^\frac{1}{\tilde{\nu} \tilde{z}},\tau L^{-z},\tau L^{-z_u}),
\end{equation}
in which $Y$ is an arbitrary operator, $s$ is the exponent related to $Y$, $\delta\equiv q-q_{\rm c}$ is the distance to the critical point, $L$ is the lattice size, and $\tilde{z}$ is the dynamic exponent, which can be $z$ or $z_u$, or their combination, depending on the operator $Y$ and the dynamic process, similarly, $\tilde{\nu}$ can be $\nu$ or $\nu'$ or \revone{combination of both, $f$ is the scaling function.} For the three kinds of initial states introduced above, the initial state information does not appear explicitly in Eq.~(\ref{eq:operator}), since all of them are the fixed points of the initial states. In contrast, for other initial states with finite initial order parameter and initial correlations, these initial conditions should be included in the scaling form~\cite{Janssen1989,Lizb1995prl,Zhengb1996prl,Albano2011iop,Yins2014prb,Yins2014pre,Shu2017prb,Shu2020prb,Zhong2021prb}.

If $z_u=z$, Eq.~(\ref{eq:operator}) recovers the usual single-length-scale relaxation scaling theory, in which, for instance, at the critical point, i.e., $\delta=0$, for a saturated initial state the order parameter scales as $M^2=\tau^{-2\beta/\nu z_u}f({\tau L^{-z_u}})$~\cite{Lizb1995prl,Albano2011iop,Shu2021prl}, while for a disordered initial state $M^2=L^{-d}\tau^{d/z_u-2\beta/\nu z_u}f({\tau L^{-z_u}})$ in which the factor $L^{-d}$ stems from the random distribution of the initial state~\cite{Albano2011iop,Shu2021prl}. For both cases, in the long-time limit, \revone{the scaling form recovers the equilibrium case, namely, }$M^2\propto L^{-2\beta/\nu}$~\cite{Albano2011iop,Shu2021prl}.

In contrast, at the critical point of $J$-$Q_3$ model, a \emph{dual dynamic scaling} appears~\cite{Shu2021prl}. Specifically, from the saturated VBS initial state, $D^2$ is controlled by $\xi$ and obeys the scaling form
\begin{equation}
\label{eq:operator1}
D^2(\tau,L)=\tau^{-\frac{\beta}{\nu z_u}}f(\tau L^{-z_u}),
\end{equation}
while $M^2$ is controlled by $\xi'$ and obeys the scaling form
\begin{equation}
\label{eq:operator2}
M^2=L^{-d}\tau^{\frac{d}{z}-\frac{2\beta}{\nu z}}f({\tau L^{-z}}).
\end{equation}
As a dual case, from the saturated AFM initial state, $M^2$ is controlled by $\xi$ and obeys the scaling form
\begin{equation}
\label{eq:operator3}
M^2(\tau,L)=\tau^{-\frac{\beta}{\nu z_u}}f(\tau L^{-z_u}),
\end{equation}
while $D^2$ is controlled by $\xi'$ and obeys the scaling form
\begin{equation}
\label{eq:operator4}
D^2=L^{-d}\tau^{\frac{d}{z}-\frac{2\beta}{\nu z}}f({\tau L^{-z}}).
\end{equation}
In addition, for the disordered initial state, which is \revone{equivalent} to the N\'{e}el and the VBS phase, both $D^2$ and $M^2$ are controlled by $\xi'$ and obey similar scaling forms
\begin{equation}
\label{eq:operator5}
P^2=L^{-d}\tau^{\frac{d}{z}-\frac{2\beta}{\nu z}}f({\tau L^{-z}}),
\end{equation}
with $P$ represents $D$ and $M$.
% and
% \begin{equation}
% \label{eq:operator6}
% M^2=L^{-d}\tau^{\frac{d}{z}-\frac{2\beta}{\nu z}}f({\tau L^{-z}}).
% \end{equation}
In the long-time limit, all these equations tend to the same equilibrium form, $D^2\sim M^2\sim L^{-2 \beta/\nu}$~\cite{Sandvik2007prl}. These scaling forms demonstrate a remarkable \emph{dual dynamic scaling} that the dynamic scaling behaviors change to their dual partners correspondingly when the initial states turn to their dual counterpart. This dual dynamic scaling reflects the equivalence between the N\'{e}el order and the VBS order at the critical point, and can thus be regarded as the noneqilibrium incarnation of the equilibrium emergent $\mathbb{SO}(5)$ symmetry.

\begin{figure}[tbp]
  \centering
  \includegraphics[width=\linewidth,clip]{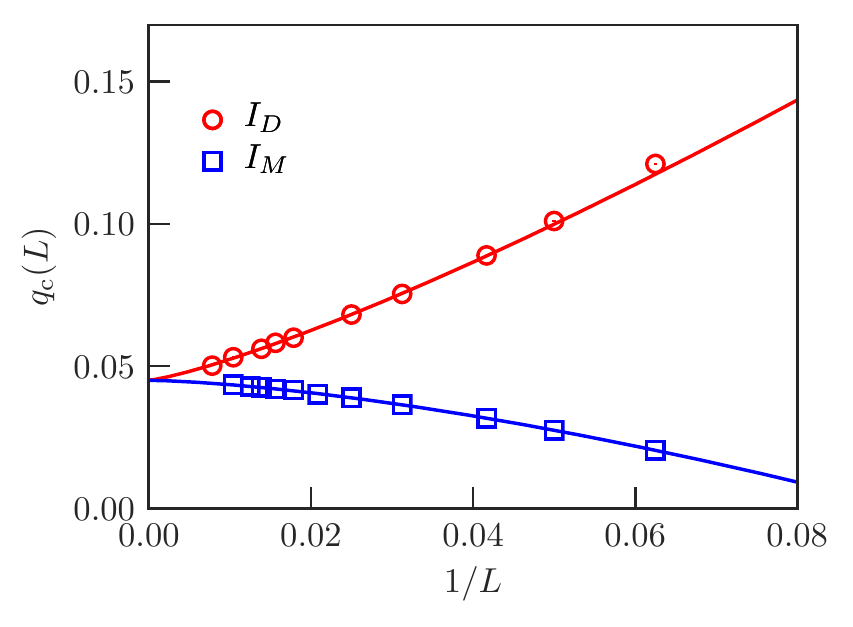}
  \vskip-3mm
  \caption{Estimation of the critical point via the dynamic scaling of the sign function of the order parameter $I_{D(M)}$. The crossing points of curves of $I_{D(M)}$ versus $q$ for $L$ and $2L$ and fixed $\tau L^{-1}=1/4$ converge to the critical point as $L\rightarrow\infty$, giving $q_{\rm c}=0.0449(7)$ from $I_{D}$ and $q_{\rm c}=0.0453(5)$ from $I_{M}$. The solid lines indicate fits of the form $q_{\rm c}(L)=q_{\rm c}+aL^{-\omega}$.
  }
  \label{fig:qcl}
\end{figure}
\begin{figure}[tbp]
\centering
  \includegraphics[width=\linewidth,clip]{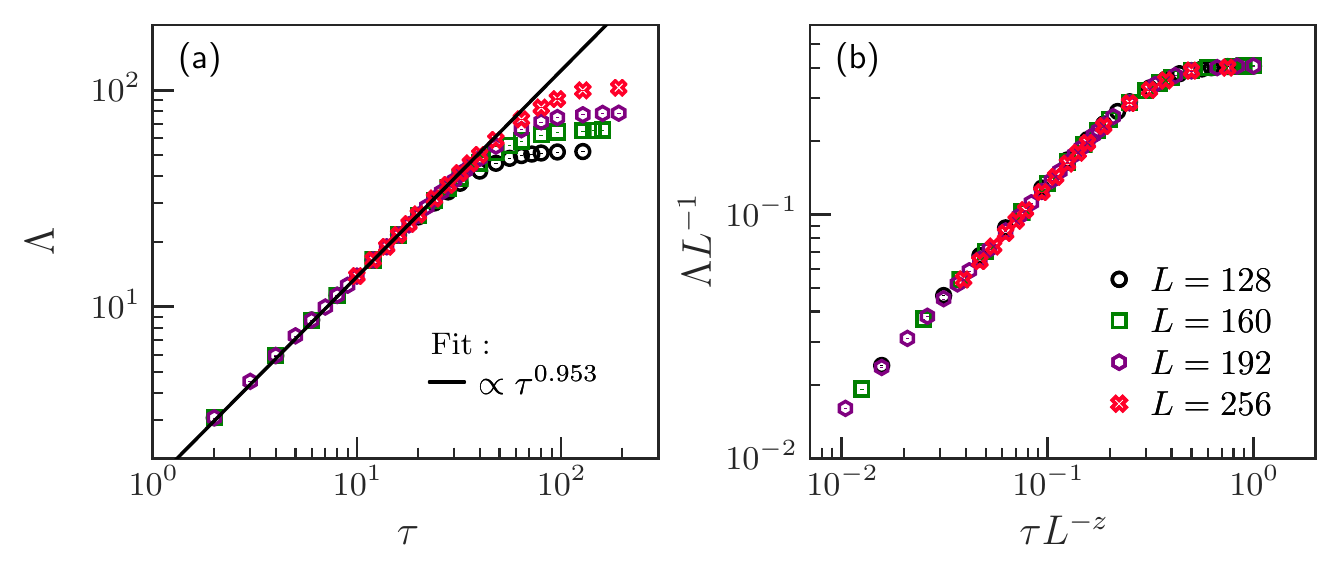}
  \vskip-3mm
  \caption{Evolution of the distance between two spinons $\Lambda$ at the critical point with the initial state set as the one with a triplet embedded in the VBS background. The curves for different $L$ before and after rescaled are shown in (a) and (b), respectively. The solid line is a power-law fit that gives an exponent of $0.953$, close to $1/z$.
  }
  \label{fig:conlength}
\end{figure}

\section{\label{criticalpoint}Estimation of the critical point}

In this section, we employ the nonequilibrium scaling to estimate the critical point of the $J$-$Q_2$ model~(\ref{eq:hamiltonian}). In equilibrium, for usual criticality with single length scale, any arbitrary dimensionless quantity $A$ scales as $A=f(\delta L^{1/\nu})$. Accordingly, for different system sizes, curves of $A$ versus $\delta$ will cross at the critical point \revone{if the scaling corrections are neglected.}
% if the scaling corrections are not taken into account.
One can use this scaling feature of dimensionless quantities to estimate the critical point~\cite{Sandvik2010review}. For the DQCP with two length scales, in general $A=f(\delta L^{1/\tilde{\nu}})$, \revone{in which the choice of $\tilde{\nu}$ depends on the quantity} . For example, \revtwo{in equilibrium,} when $A$ is the Binder ratio of the order parameter, $\tilde{\nu}=\nu$; when $A$ is the ratio of the distance between two spinons to the lattice, $\tilde{\nu}=\nu'$~\cite{Sandvik2016science}. Both can be employed to determine the critical point~\cite{Sandvik2016science}.

In generalizing the scaling form of $A$ to the nonequilibrium case, according to Eq.~(\ref{eq:operator}), one find that in general $A=f(\tau L^{-z},\tau L^{-z_u},\delta L^{1/\tilde{\nu}})$, in which there are two additional time-dependent variables: $\tau L^{-z}$ and $\tau L^{-z_u}$. Without a priori wisdom, one does not know which one dominates. In the $J$-$Q_3$ model, it was found that the sign function of the VBS order parameter from the VBS initial state, $I_{D}\equiv\langle {\rm sgn}(D)\rangle$ is dominated by $\tau L^{-z}$ and so does the sign function of the N\'{e}el order parameter from the AFM initial state, $I_{M}\equiv\langle {\rm sgn}(M)\rangle$~\cite{Shu2021prl}.

To verify the universality of the dynamic properties of $I_{D(M)}$ and determine the critical point of the $J$-$Q_2$ model, here we calculate $I_D$ and $I_M$ versus $\delta$ for various lattice sizes with fixed $\tau L^{-z}=1/4$ from the VBS and the AFM initial states, respectively.
\revtwo{
  Note that in principle, different values of the ratio $\tau L^{-z}$ should not deliver distinguishable results of $q_{\rm c}$ as long as $L$ is large enough.}

From Fig.~\ref{fig:qcl}, one finds that the \revone{size dependence of the} crossing points of $I_D$ versus $q$ for $L$ and $2L$ decrease monotonously as $L$ increases and converge to a point $q=q_{\rm c}=0.0449(7)$ \revone{in the thermodynamic limit}, which is close to the known results of the critical point~\cite{Sandvik2010prl,Sandvik2016science,Sandvik2020iop}. Moreover, Fig.~\ref{fig:qcl} also shows that the crossing points of the curves of $I_M$ versus $q$ for $L$ and $2L$ increases as $L$ increases and converge to $q=q_{\rm c}=0.0453(5)$, which is almost \revtwo{identical} as that in $I_D$ case within the error bar. These results not only determine the critical point accurately, but also demonstrate that both $I_D$ and $I_M$ are dominated by $\tau L^{-z}$ in the relaxation process from \revtwo{their respective} ordered initial states, similar to the case of the $J$-$Q_3$ model. In the following calculations, we use $q_{\rm c}=0.045$, which is closed to the average value of $q_{\rm c}$ obtained by $I_D$ and $I_M$.

To further examine the value of the critical point, we calculate the dynamic scaling of the spinon confinement length at the critical point $q_{\rm c}=0.045$. By setting the initial state as that with a triplet in the VBS background, we calculate the averaged distance between two unpaired spinons, $\Lambda$, which is proportional to the confinement length $\xi'$~\cite{Sandvik2016science}. Figure~\ref{fig:conlength} (a) shows that in the short-time stage, $\Lambda\propto \tau^{0.953}$ with the exponent close to $1$. Moreover, Fig.~\ref{fig:conlength} (b) shows that the rescaled curves of $\Lambda L^{-1}$ versus $\tau L^{-z}$ collapse onto each other well. These results not only confirm the value of the critical point, but also demonstrate that the scaling relation \revone{$\xi'\propto \tau^{1/z}$} is universal in both the $J$-$Q_2$ and the $J$-$Q_3$ model. 

However, different from the usual criticality with \revtwo{single} length scale, in the DQCP with two length scales, not every dimensionless variable exhibits scaling properties controlled by the same scaling variable. In Sec.~\ref{dualscalinga}, we will show that the dynamics of the Binder ratios of the order parameters can be dominated by both $\tau L^{-z}$ and $\tau L^{-z_u}$, depending on the initial states.

\begin{figure*}[t]
\centering
  \includegraphics[width=\linewidth,clip]{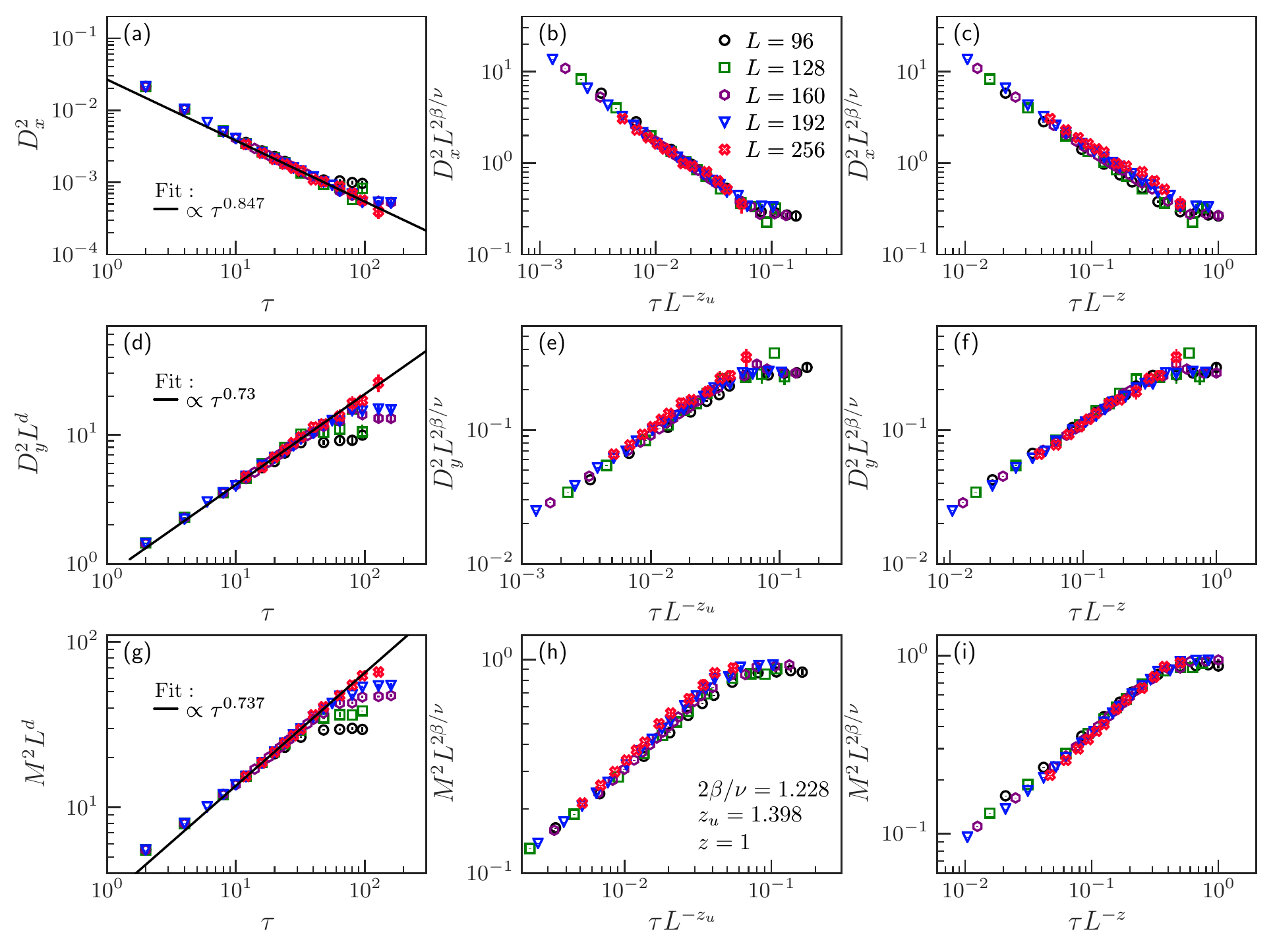}
  \vskip-3mm
  \caption{Relaxation dynamics of the order parameters with the saturated VBS initial state. Evolution of $D_x^2$, $D_y^2$, and $M^2$ for various lattice sizes $L$ indicated are shown in top (a)-(c), middle (d)-(f), and bottom (g)-(i) rows, respectively. In the left column (a),(d),(g), curves of the order parameter versus $\tau$ are fitted in the short-time stage by the power function. In addition, curves after rescaled are shown in the middle (b),(e),(h) and right (c),(f),(i) columns. For both columns, $D_x^2$, $D_y^2$, and $M^2$ are rescaled as $D_x^2L^{2\beta/\nu}$,$D_y^2L^{2\beta/\nu}$ and $M^2L^{2\beta/\nu}$, respectively. To compare, $\tau$ is rescaled according to $\tau L^{-z_u}$ and $\tau L^{-z}$ in the middle (b),(e),(h) and right (c),(f),(i) columns, respectively.
  }
  \label{fig:vbsors}
\end{figure*}

\section{\label{dualscaling}Dual dynamic scaling at the critical point}

In this section, we will explore the imaginary-time relaxation dynamics at the critical point from different initial states. We will show that the dual dynamic scaling is a universal behavior in the relaxation dynamics of the DQCP since it also appears at the critical point of the $J$-$Q_2$ model.

\subsection{\label{dualscalinga}Dynamics with the VBS initial state}

\begin{figure*}[htb]
\centering
  \includegraphics[width=\linewidth,clip]{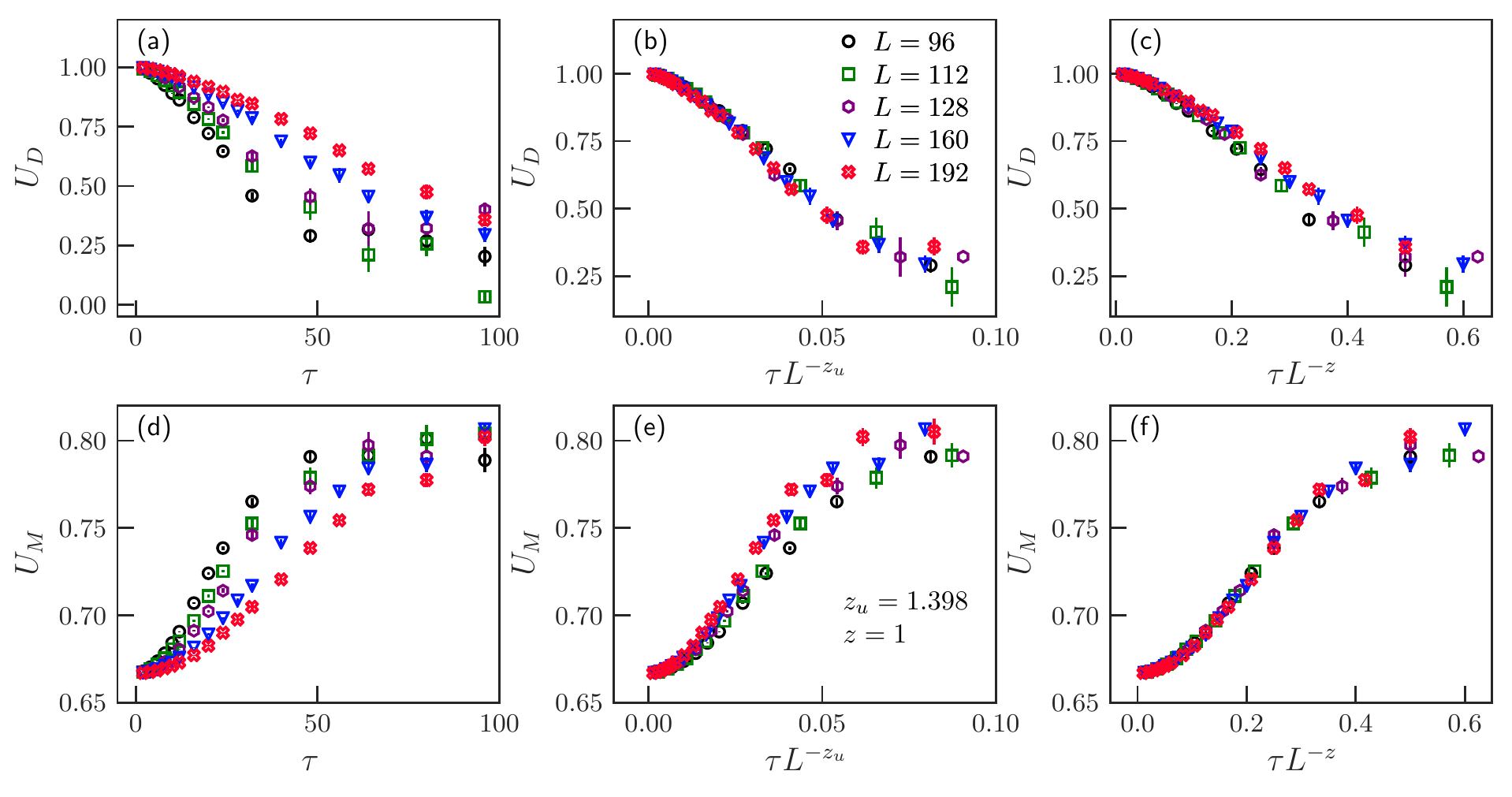}
  \vskip-3mm
  \caption{Relaxation dynamics of the Binder ratios with the VBS initial state. Evolutions of $U_D$ and $U_M$ for various lattice sizes $L$ indicated are shown in top (a)-(c) and bottom (d)-(f) rows, respectively. For comparison, $\tau$ is rescaled according to $\tau L^{-z_u}$ in the middle column (b),(e); while $\tau$ is rescaled according to $\tau L^{-z}$ in the right column (c),(f).
  }
  \label{fig:binder}
\end{figure*}

The VBS order breaks the $\mathbb{Z}_4$ discrete symmetry. The initial VBS state is chosen as that with horizontal dimers occupying every other bond as shown in Fig.~\ref{fig:quench}. Accordingly, at $\tau=0$, $D_x$ has a saturated value \revtwo{$D_{x,0}=3/8$}, while $D_y=0$.

For the VBS order parameter $D_x$, we find from Fig.~\ref{fig:vbsors} (a) that it relaxes according to $D_x^2\propto \tau^{-0.847}$ in the short-time stage. This exponent is close to $2\beta/\nu z_u \simeq 0.881$ rather than $2\beta/\nu z \simeq 1.228$, in which $z_{u}=z\nu'/\nu \simeq 1.398$. \revtwo{Here, we use $\nu/\nu'\simeq 0.715$ obtained from direct fitting of the exponent ratio in Ref.~\cite{Sandvik2016science}}.
Therefore, in general, the full scaling form of $D_x^2$ in the whole relaxation process should be $D_x^2=\tau^{-2\beta/\nu z_u}f(\tau L^{-z_u},\tau L^{-z})$. Moreover, previous studies showed that in equilibrium $D^2\propto L^{-2\beta/\nu}$~\cite{Sandvik2007prl,Sandvik2010prl}. Thus one can infer that the full scaling form of $D_x^2$ should also be $D_x^2=L^{-2\beta/\nu}g(\tau L^{-z_u},\tau L^{-z})$, \revone{in which $g$ is another scaling function}. These two full scaling forms should be consistent with each other. This constraint indicates that the dominant scaling variable in the scaling function is $\tau L^{-z_u}$ rather than $\tau L^{-z}$, so that the scaling form can be converted from $D_x^2=\tau^{-2\beta/\nu z_u}f(\tau L^{-z_u})$ to $D_x^2=L^{-2\beta/\nu}g(\tau L^{-z_u})$ by recognizing the relation $f(\tau L^{-z_u})=(\tau L^{-z_u})^{2\beta/\nu z_u}g(\tau L^{-z_u})$. To verify the scaling theory, in Figs.~\ref{fig:vbsors} (b) and (c), we plot the rescaled curves of $D_x^2L^{2\beta/\nu}$ versus $\tau L^{-z_u}$ and compare them with those of $D_x^2L^{2\beta/\nu}$ versus $\tau L^{-z}$.
\revone{It is clear that with $\tau$ scaled with $L^{-z_u}$, the curves collapse better than with $L^{-z}$, indicating that \revtwo{it is $\tau L^{-z_u}$, or $\xi$,  that governs the scaling form.
  }}
These results show that the scaling form of $D_x^2$ from the VBS initial state also satisfied Eq.~(\ref{eq:operator1}) found in the $J$-$Q_3$ model~\cite{Shu2021prl}, confirming the universality of the scaling of $D_x^2$ with the horizontal VBS initial state in the N\'{e}el-VBS transition.

Note that here one may argue that $D_x^2$ can also be expressed as $D_x^2(\tau, L)\propto \tau^{- 2\beta/\nu' z}$. In this case, $2\beta/\nu'\simeq 0.881$ is smaller than $1$. Accordingly, the scaling law $2\beta/\nu'-1=\eta'$ gives a negative anomalous dimension $\eta'$, which would imply a nonunitary theory~\cite{Nahum2015prx,Ferrara1974prd,Mack1977}. Alternatively, to satisfy the unitarity bound of the critical point~\cite{Ferrara1974prd,Mack1977}, we choose to adopt $D_x^2(\tau, L)\propto \tau^{- 2\beta/\nu z_u}$ in which ${2 \beta/\nu}$ keeps intact while an additional dynamic exponent $z_u$ is introduced.

From the perspective of the N\'{e}el order, the saturated VBS state \revone{plays a similar role to} a disordered state, since both the VBS state and the disordered state keep the spin rotation symmetry and have vanishing correlation length. For the N\'{e}el order parameter $M$, we find from Fig.~~\ref{fig:vbsors} (g) that it relaxes according to $M^2\propto L^{-d} \tau^{0.737}$ in the short-time stage. The exponent $0.737$ is close to $(d/z-2\beta/\nu z)=0.772$. Thus the general full scaling form characterizing the whole relaxation process should be $M^2=L^{-d}\tau^{d/z-2\beta/\nu z}f(\tau L^{-z_u},\tau L^{-z})$. Similar to $D_{x}^2$, the full scaling form can also be expressed as $M^2=L^{-2 \beta/\nu}g(\tau L^{-z_u},\tau L^{-z})$, since in equilibrium $M^2\propto L^{-2\beta/\nu}$. The consistence between these two full scaling forms dictates that the dominant scaling variable in the scaling function is $\tau L^{-z}$ rather than $\tau L^{-z_u}$,
% because if so
\revone{such that} the scaling form can be converted from $M^2=L^{-d}\tau^{d/z-2\beta/\nu z}f(\tau L^{-z})$ to $M^2=L^{-2\beta/\nu}g(\tau L^{-z})$ by recognizing the relation \revtwo{$f(\tau L^{-z})=(\tau L^{-z})^{-d/z+2\beta/\nu z}g(\tau L^{-z})$}. Moreover, by comparing the rescaled curves of $M^2L^{2\beta/\nu}$ versus $\tau L^{-z_u}$ and $\tau L^{-z}$, one finds from Figs.~\ref{fig:vbsors} (h)-(i) that the rescaled curves collapse onto each other with $\tau L^{-z}$, while deviate from each other with $\tau L^{-z_u}$. These results show that in the $J$-$Q_2$ model the dynamics of $M^2$ from the VBS initial state is controlled by the confinement length scale $\xi'$ and satisfied Eq.~(\ref{eq:operator2}), same as the case of the $J$-$Q_3$ model~\cite{Shu2021prl}, confirming the universality of the dynamic critical behavior of $M^2$ in DQCP.

In addition, we calculate the VBS order parameter in the vertical direction $D_y$. From Fig.~\ref{fig:vbsors} (d) we find that it relaxes according to $D_y^2\propto L^{-d} \tau^{0.73}$ in the short-time stage. The exponent is close to $(d/z-2\beta/\nu z)$, in analogy to the case of $M^2$. Similar analyses give that the full scaling forms should be $D_y^2=L^{-d}\tau^{d/z-2\beta/\nu z}f(\tau L^{-z})$ and $D_y^2=L^{-2\beta/\nu}g(\tau L^{-z})$. Moreover, by comparing the rescaled curves of $D_y^2L^{2\beta/\nu}$ versus $\tau L^{-z_u}$ and $\tau L^{-z}$, one finds that the rescaled curves collapse better with $\tau L^{-z}$, as shown in Figs.~\ref{fig:vbsors} (e)-(f).
From the above results, one can see that in the $J$-$Q_2$ model, with the horizontal VBS initial state, the dynamics of $D_y^2$ is also controlled by the confinement length scale, same as the case of $M^2$. Such results illustrate the equivalence under the rotation between $M$ and $D_y$ \revtwo{when the initial state is saturated in $D_x$}.

To further reveal the dynamic scaling behaviors from the VBS initial state, we also study the Binder ratios for the VBS and N\'{e}el order parameters, defined as $U_D\equiv\frac{3}{2}-\frac{\langle D_x^4\rangle}{2\langle D_x^2\rangle^2}$ and $U_M\equiv\frac{3}{2}-\frac{\langle M^4\rangle}{2\langle M^2\rangle^2}$, respectively. As a dimensionless quantity, the general scaling forms are $U_{D(M)}=f_{D(M)}(\tau L^{-z_u},\tau L^{-z})$. By comparing $U_{D(M)}$ versus $\tau L^{-z_u}$ and $\tau L^{-z}$ in Fig.~\ref{fig:binder}, one finds that $U_D$ is controlled by the usual length scale $\xi$ and its dynamics obeys $U_{D}=f_{D}(\tau L^{-z_u})$, while $U_M$ is controlled by the confinement length scale $\xi'$ and its dynamics obeys $U_{M}=f_{M}(\tau L^{-z})$. These scaling properties are also consistent with the results that $D_x^2$ and $M^2$ are dominated by $\tau L^{-z_u}$ and $\tau L^{-z}$, respectively.
\revone{Therefore, special attention should be paid when one employs the short-time dynamics of the Binder ratios $U_{D}$ and $U_{M}$ to determine the critical properties in DQCP with two length scales, although this method is often used in usual LGW phase transitions~\cite{Lizb1995prl}.} 
Combining with the scaling properties of the sign function $I_{D(M)}$, we find that in the nonequilibrium dynamics of DQCP with two length scales, dimensionless quantities can be controlled by different length scales, depending on the physical quantities themselves and the initial states applied.

\subsection{Dynamics with the AFM initial state}

\begin{figure*}[t]
\centering
  \includegraphics[width=\linewidth,clip]{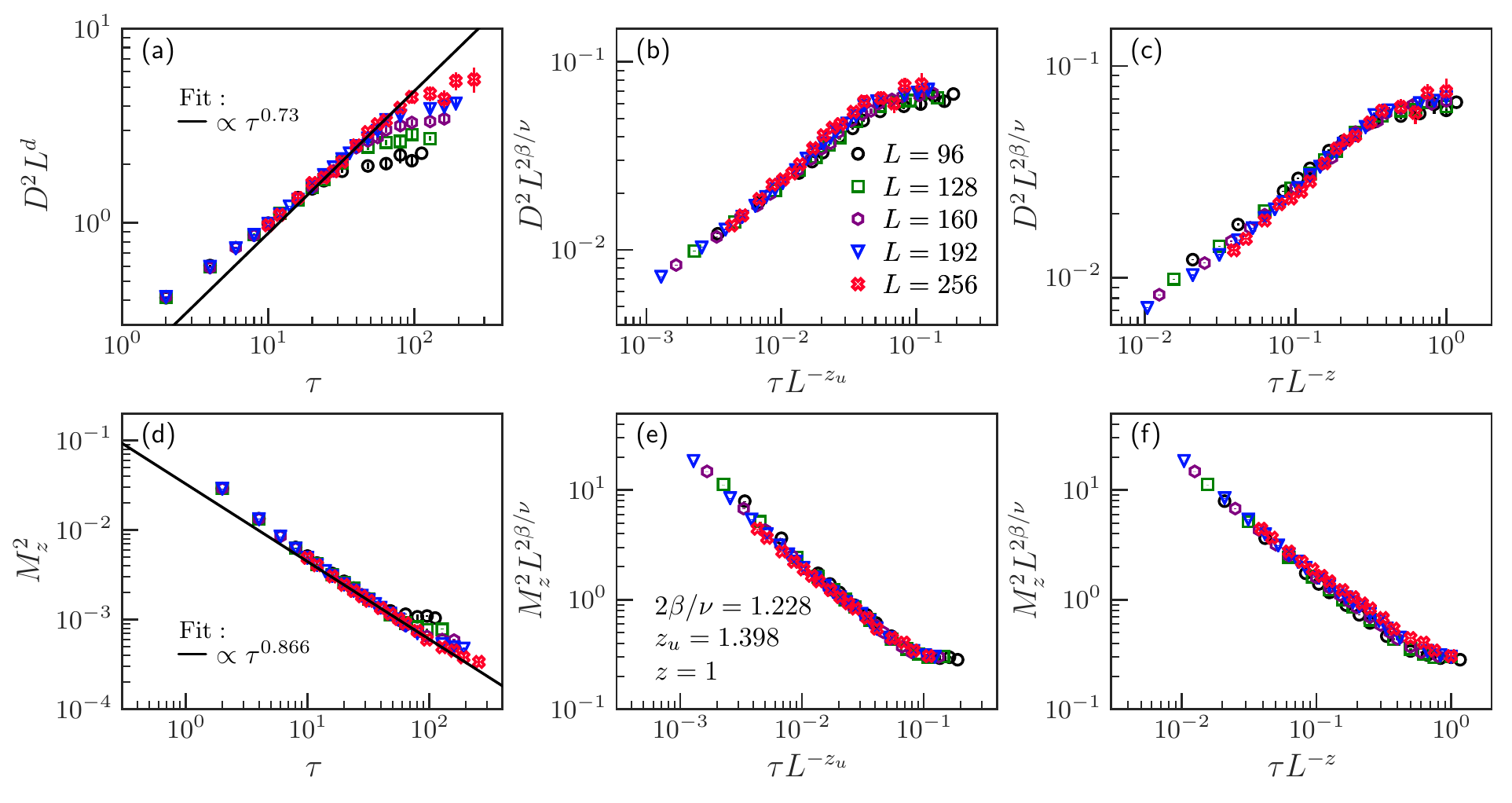}
  \vskip-3mm
  \caption{Relaxation dynamics of order parameters with the AFM initial state. Evolutions of $D^2$ and $M_z^2$ for various lattice sizes $L$ indicated are plotted in top (a)-(c) and bottom (d)-(f) rows, respectively. In the left column (a),(d), curves of the order parameter versus $\tau$ are fitted in the universal short-time stage by the power function. In addition, curves after rescaled are shown in the middle (b),(e) and right (c),(f) columns. For both columns, $D^2$ and $M^2$ is rescaled as $D^2L^{2\beta/\nu}$ and $M_z^2L^{2\beta/\nu}$, respectively. In contrast, $\tau$ is rescaled according to $\tau L^{-z_u}$ and $\tau L^{-z}$in the middle (b),(e) and right (c),(f) columns, respectively.
  }
  \label{fig:afors}
\end{figure*}

The AFM state breaks the spin $\mathbb{SO}(3)$ rotation symmetry. We choose the initial state as the one with saturated component in the $z$-direction, as shown in Fig.~\ref{fig:quench}.

In analogy to the situation of the N\'{e}el order with the saturated VBS initial state, from the perspective of the VBS order, the saturated AFM state \revone{plays a similar role} to a disordered state, since both of them contribute zero VBS order parameter and vanishing correlation length. We find in Fig.~\ref{fig:afors} (a) that the VBS order parameter evolves according to $D^2\propto L^{-d} \tau^{0.73}$ in the universal short-time stage. \revone{Here, $D^2=D_x^2+D_y^2$.} This exponent is close to $(d/z-2\beta/\nu z)$. By considering the equilibrium scaling $D^2\propto L^{-2\beta/\nu}$ simultaneously~\cite{Sandvik2007prl,Sandvik2010prl}, one finds that the confinement length dominates the dynamics and the scaling form of $D^2$ is $D^2=L^{-d}\tau^{d/z-2\beta/\nu z}f(\tau L^{-z})$. This scaling form can be converted to $D^2=L^{-2\beta/\nu}g(\tau L^{-z})$ by substituting \revtwo{$f(\tau L^{-z})=(\tau L^{-z})^{-d/z+2\beta/\nu z}g(\tau L^{-z})$} into the former equation. Moreover, by comparing the rescaled curves of $D^2L^{2\beta/\nu}$ versus $\tau L^{-z_u}$ and $\tau L^{-z}$, as shown in Figs.~\ref{fig:afors} (b)-(c), one finds that the rescaled curves collapse better with $\tau L^{-z}$. These results show that in the $J$-$Q_2$ model the dynamics of $D^2$ from the AFM initial state is controlled by the confinement length scale and satisfied Eq.~(\ref{eq:operator4}). These results are the same as the case of the $J$-$Q_3$ model~\cite{Shu2021prl}, confirming the universality of the dynamic critical behavior of $D^2$ in DQCP.

\begin{figure*}[htb]
\centering
  \includegraphics[width=\linewidth,clip]{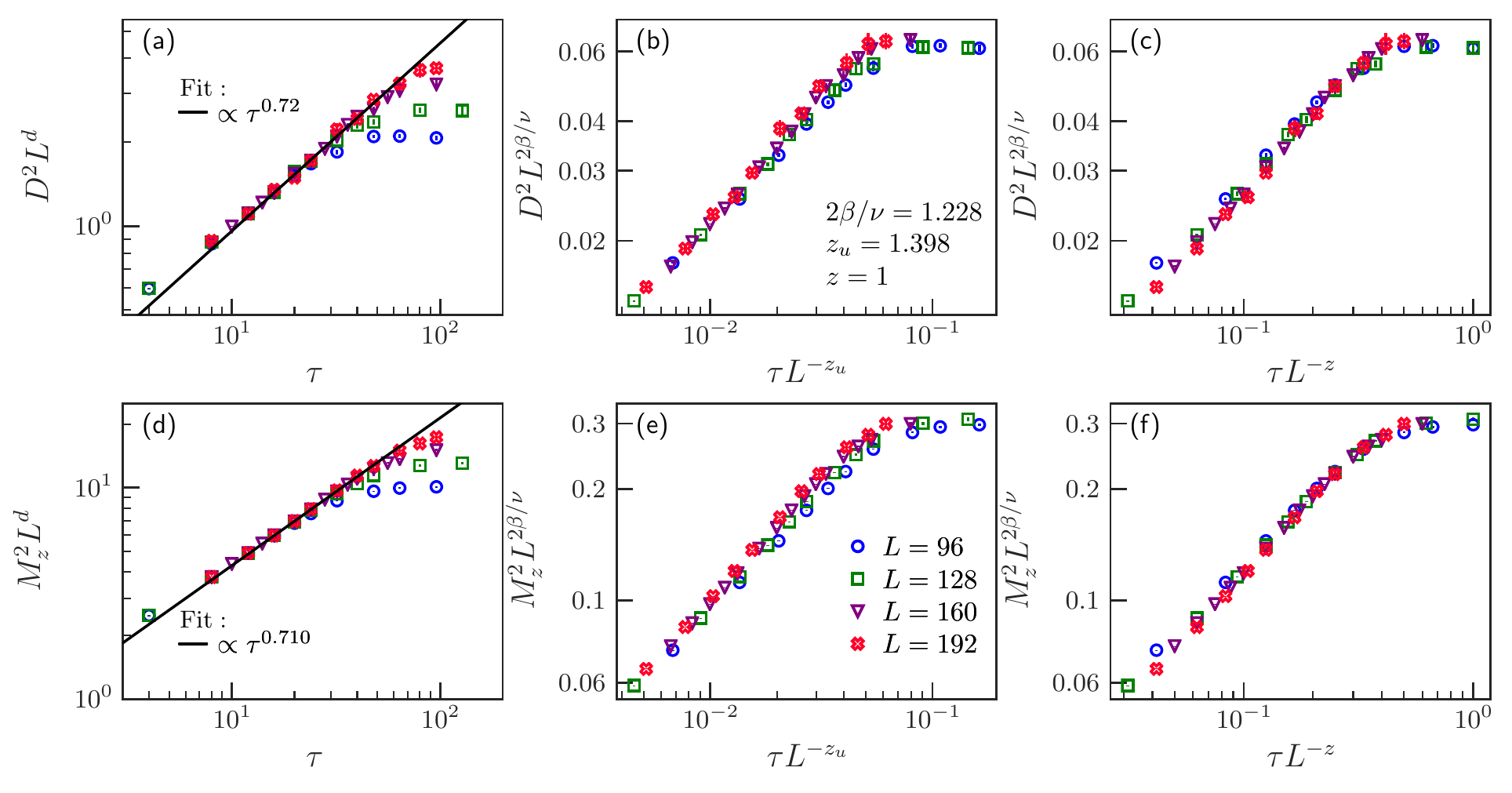}
  \vskip-3mm
  \caption{Relaxation dynamics of order parameters with the disordered initial state. Evolutions of $D^2$ and $M^2$ for various lattice sizes $L$ indicated are plotted in top (a)-(c) and bottom (d)-(f) rows, respectively. In the left column (a),(d), curves of the order parameter versus $\tau$ are fitted in the short-time stage by the power function. In addition, curves after rescaled are shown in the middle (b),(e) and right (c),(f) columns. For both columns, $D^2$ and $M^2$ is rescaled as $D^2L^{2\beta/\nu}$ and $M_z^2L^{2\beta/\nu}$, respectively. In contrast, $\tau$ is rescaled according to $\tau L^{-z_u}$ and in the middle (b),(e) and right (c),(f) columns, respectively.
  }
  \label{pmors}
\end{figure*}

For the N\'{e}el order parameter, similar to the VBS order parameter from the saturated VBS initial state, it decays with $\tau$ as $M_z^2\propto \tau^{-0.866}$ with the exponent close to $2\beta/\nu z_u$ as shown in Fig.~\ref{fig:afors} (d). Combining with the equilibrium finite-size scaling $M_z^2\propto L^{-2\beta/\nu}$~\cite{Sandvik2007prl,Sandvik2010prl}, one deduces that the full scaling form should be $M_z^2=\tau^{-2 \beta/\nu z_u}f(\tau L^{-z_u})$, or equivalently, $M_z^2=L^{-2 \beta/\nu}g(\tau L^{-z_u})$. As seen in Figs.~\ref{fig:afors} (e)-(f), by rescaling $M_z^2$ and $\tau$ with $L^{2\beta/\nu}$ and $L^{-z_u}$, we find the rescaled curves match with each other. However, when $\tau$ is rescaled with $L^{-z}$, the rescaled curves deviates. These results confirm that the dynamics of $M_z^2$ with the $z$-component AFM initial state is controlled by the usual correlation length scale and satisfied Eq.~(\ref{eq:operator3}). These results are also the same as the case of the $J$-$Q_3$ model~\cite{Shu2021prl}.

\subsection{Dynamics with the disordered initial state}

The disordered initial state can be prepared in very high temperature region.
\revtwo{From the perspective of the disordered state, either the N\'{e}el or the VBS order looks equivalent.}
% From either the VBS side or the N\'{e}el order, the disordered state looks equivalent.
Accordingly, it is expected that both $D^2$ and $M^2$ should satisfy the same scaling form: $P^2=L^{-d}\tau^{d/z-2\beta/\nu z}f(\tau L^{-z})$, in which $P$ represents $D$ and $M$. This dynamic scaling is similar to the cases of $D^2$ with the AFM initial state and $M^2$ with the VBS initial state. To verify these scaling properties, in Figs.~\ref{pmors} (a) and (d), we plot the evolution of $D^2$ and $M^2$. We find that in the universal short-time stage $P^2\propto L^{-d}\tau^{d/z-2\beta/\nu z}$. In addition, by noting that $P^2=L^{-d}\tau^{d/z-2\beta/\nu z}f(\tau L^{-z})$ can be transformed to $P^2=L^{-2\beta/\nu z}f(\tau L^{-z})$, we compare the curves of $P^2L^{2 \beta/\nu}$ versus $\tau L^{-z}$ and $\tau L^{-z_u}$, respectively. From Figs.~\ref{pmors} (b)-(c) and (e)-(f), one finds that with $\tau L^{-z}$, the curves collapse better, verifying the full scaling forms mentioned above. Moreover, these results demonstrate that both $M^2$ and $D^2$ are controlled by the confinement length $\xi'$. The same scaling properties are also found in the $J$-$Q_3$ model as shown in Eq.~(\ref{eq:operator5}). Therefore one concludes that these scaling properties are universal as well.

\subsection{Dual dynamic scaling}

The appearance of the emergent symmetry is a characteristic critical property of the DQCP~\cite{Senthil2004science,Sachdev2004prb,Sandvik2007prl,Sandvik2009prb1,Nahum2015prl,Nahum2019prl,Mans2019prl}. For \revone{the class of the $\mathbb{SU}(2)$} $J$-$Q$ model, the emergent symmetry at the critical point is the $\mathbb{SO}(5)$ symmetry, which describes the rotation symmetry between the components of the superspin $\mathcal{S}=(M_x,M_y,M_z,D_x,D_y)$, including the components of the N\'{e}el and the VBS order parameters~\cite{Senthil2004science,Sachdev2004prb,Sandvik2007prl,Sandvik2009prb1,Nahum2015prl,Nahum2019prl}. The emergent symmetry is broken in both ordered phases. In particular, along with the appearance of the discrete $\mathbb{Z}_4$ symmetry, scaling properties with two length scales arise on the VBS side.

%This continuous emergent symmetry is associated with the conservation of the density of the monopoles at the critical point, indicating the critical phenomenon is described by the noncompact quantum electrodynamics in (2+1)D. Accordingly, the fugacity of the monopole arises as a dangerously irrelevant scaling variable, which is irrelevant at the critical point, but in the VBS phase it becomes relevant along with the breakdown of the continuous symmetry to the $\mathbb{Z}_4$ discrete symmetry. According to the critical theory with the dangerously irrelevant scaling variable, in the VBS side, there should be two divergent length scales in controlling the critical properties.

In the imaginary-time relaxation dynamics, the above numerical results for the $J$-$Q_2$ model (\ref{eq:hamiltonian}) show a remarkable dual dynamic scaling behavior: the scaling forms of the order parameters exchange as the initial state is rotated in the superspin space, similar to the results in the $J$-$Q_3$ model~\cite{Shu2021prl}. Specifically, when the initial state is rotated from the VBS state to the AFM state, the dynamic scaling form of the VBS (N\'{e}el) order parameter changes to that of the N\'{e}el (VBS) order parameter, and vice versa. Additionally, for the disordered initial state which keeps invariant under the rotation between the superspin components, both the VBS and the N\'{e}el order parameters show similar scaling behaviors. Comparing with the equilibrium emergent symmetry, we find that the dual dynamic scaling also reflects the rotation symmetry between the superspin components. Moreover, these results also show that effects induced by the interplay between two length scales are naturally included in the dual dynamic scaling.

A prominent question is why the relaxation processes of different superspin components can be \revone{governed} by different length scales, albeit their equilibrium finite-size scaling forms are the same. By inspecting
\revtwo{the relaxation behaviors under different initial conditions,}
% the initial condition and the following relaxation,
one finds that for the dominant component with a saturated initial value, i.e., the component accordant with the saturated initial component, its relaxation dynamics is controlled by the usual correlation length; while for the complementary component with zero initial value, i.e., the component orthogonal to the saturated initial component, its relaxation dynamics is controlled by the the confinement length. For the \revone{dominant component}, the relaxation dynamics is associated with the local fluctuations whose characteristic length scale is measured by the usual correlation length $\xi$. For \revone{the complementary component}, the average value of the component keeps zero in the relaxation process due to the symmetry of the Hamiltonian, while its average squared value in the short-time stage is proportional to $L^{-d}$ \revtwo{at a given evolution time $\tau$}. Since this average squared value is directly related to the lattice size, one can infer that its dynamics is associated with the global fluctuations, which is related to the topological properties of the system.
\revtwo{Different from the situation in the usual critical point where both the local and the global fluctuations have the same characterized length scale $\xi$~\cite{Albano2011iop,Shu2021prl}, in the DQCP 
}
% In the usual critical point, both global fluctuations and the local fluctuations have the same characterized length scale $\xi$~\cite{Albano2011iop,Shu2021prl}. In contrast, in the DQCP
of the $J$-$Q$ model \revone{class}, the global fluctuations can have a different typical length scale. To be specific, the global fluctuations correspond to the excitations with the spinons living in the vortices of the VBS domain walls and the distance between these spinons is characterized by the confinement length $\xi'$~\cite{Senthil2004prb}. Accordingly, it is expected that the \revone{complementary} order parameter component with a vanishing initial value should be controlled by the confinement length $\xi'$. This argument is also supported by the findings that with a disordered initial state, \revone{all superspin components are governed by global fluctuations, so that both $M^2$ and $D^2$ are controlled by $\xi'$}.

Besides the order parameters, other quantities can also obey the dual dynamic scaling. For instance, it is shown that the sign function $I_{D(M)}$ from the VBS (N\'{e}el) saturated initial state is controlled by the confinement length $\xi'$. By rotating $D$ to $M$, the initial states and dynamic scaling forms change correspondingly. Note that here the scaling form for $I_{D}$ and $I_M$ are the same. Moreover, although $D^2$ and $M^2$ are controlled by the usual length scale $\xi$ for their respective ordered initial states, their sign functions are dictated by the global flip of the order parameter over entire lattice range. Accordingly, $I_{D(M)}$ is controlled by the confinement length.

\begin{figure*}[htb]
\centering
  \includegraphics[width=\linewidth,clip]{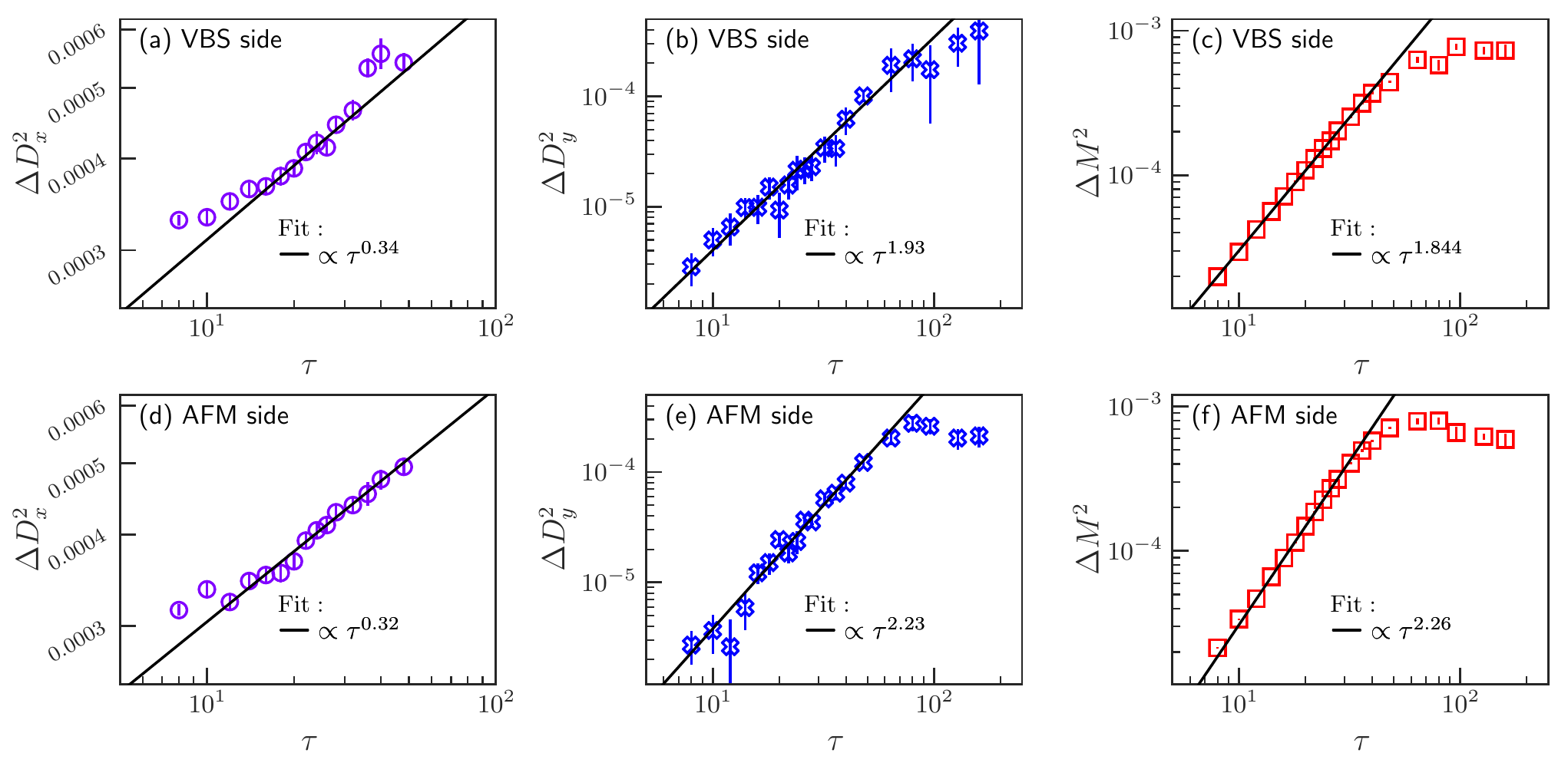}
  \vskip-3mm
  \caption{Relaxation dynamics away from the critical point with the VBS initial state. Top row: for $\delta>0$, curves of $\Delta D_x^2$, $\Delta D_y^2$ and, $\Delta M^2$ versus $\tau$ are plotted in (a), (b), and (c), respectively. Bottom row: for $\delta<0$, curves of $\Delta D_x^2$, $\Delta D_y^2$ and, $\Delta M^2$ versus $\tau$ are plotted in (d), (e), and (f), respectively. Power law fittings are implemented for all curves in the universal short-time stage. Double-logarithmic scales are used.
  }
  \label{fig:initvbsgg}
\end{figure*}

Here we discuss another intriguing puzzle on the dual dynamic scaling. According to the usual critical theory in the presence of a dangerously irrelevant scaling variable, the additional length scale $\xi'$ only play roles on one side of the critical point. Accordingly, asymmetric scaling properties arise between two sides of the critical point in equilibrium~\cite{Delamotte2015prl}.
\revone{
  In the DQCP, for the relaxation dynamics from an ordered initial state to the critical point, this selected ordered state breaks the rotation symmetry in the superspin space. Therefore, the relaxation process should encode the information of the initial ordered phase during its way down to the ground state. As a result, it is expected that with different ordered initial states, asymmetric dynamic scaling should appear rather than the dual dynamic scaling. Namely, following this argument, for the dominant order parameter, different scaling forms should be observed for the dominant order parameter on the exchange of the saturated AFM and VBS initial state. So does the complementary ones.
  However, such prediction is inconsistent with the numerical results found here and in Ref.~\cite{Shu2021prl}.
  }
  % In the nonequilibrium relaxation dynamics of the DQCP, even at the critical point, a selected ordered initial state breaks the rotation symmetry in the superspin space and bring the information of the ordered phase to the critical relaxation process.
  % Thus, it is expected that asymmetric dynamic scaling should have appeared for different initial states and the dual dynamic scaling should not have come out. But this perdition is inconsistent with the numerical results.

  To solve this puzzle, we note that in equilibrium, only the ground state contributes to the expectation value and brings about the asymmetric scaling properties. In contrast, in the universal stage of nonequilibrium process, the relaxation state is the superposition of the initial state, the ground state and the low-energy excited states. Specifically, for the AFM initial state, besides the usual spin wave excitation,
\yurong{excited states with quadrupled monopoles , which are directly connect to the VBS order, also play important roles at the critical point}~\cite{Haldane1988prl,Read1990prb,Sachdev1991prl}. Although the spin wave excitations near the AFM initial state exhibit the usual correlation length $\xi$, the excited states with quadrupled monopoles should be characterized by the confinement length scale $\xi'$. Similarly, for the VBS initial state, besides the usual spin triplet excitation, excited states with spinons staying at the vortices of the VBS domain walls \yurong{that have direct connection to the N\'{e}el order, are also important at the critical point}~\cite{Senthil2004prb}. The correlation of spin triplet excitations exhibits the usual correlation length $\xi$, while the excited states combining deconfined spinons and vortices of the VBS domain walls are characterized by the confinement length scale $\xi'$. Therefore, the relaxation state \revone{is an intertwining state of both length scales, despite of the choice of ordered initial state}. This argument may help to explain the appearance of the dual dynamic scaling under the superspin rotation.
  
\section{\label{offdualscaling}Dynamic scaling away from the critical point}
In equilibrium, when the tuning parameter $\delta$ deviates from the critical point, the emergent symmetry fades away. As the nonequilibrium incarnation of the emergent symmetry, the fate of the dual dynamic scaling in the presence of a finite $\delta$ is studied in this section.

\subsection{Off-critical-point dynamic scaling from the VBS initial state}

According to Eq.~(\ref{eq:operator}), with the saturated horizontal VBS initial state and a small $\delta$, $D_x^2$ obeys the scaling form
\begin{equation}
\label{eq:operator7}
D_x^2(\tau,\delta,L)=\tau^{-\frac{\beta}{\nu z_u}}f(\delta \tau^{\frac{1}{\tilde{\nu} \tilde{z}}},\tau L^{-z_u}),
\end{equation}
in which $\tilde{\nu}$ is $\nu$ or $\nu'$ and $\tilde{z}$ is $z$ or $z_u$. However, the specific choices are unknown. To determine them, we expand the scaling function $f$ in term of $\delta \tau^{1/\tilde{\nu} \tilde{z}}$ and obtain $\Delta D_x^2\equiv D_x^2(\tau,\delta,L)-D_x^2(\tau,L)$ up to the leading order as
\begin{equation}
\label{eq:operator8}
\Delta D_x^2(\tau,\delta)\simeq \delta\tau^{-\frac{\beta}{\nu z_u}+\frac{1}{\tilde{\nu} \tilde{z}}}.
\end{equation}
Note that in Eq.~(\ref{eq:operator8}) the dependence on $L$ is ignored since in the short-time stage $\xi$ and $\xi'$ are smaller than $L$. Similarly, $\Delta D_y^2\equiv D_y^2(\tau,\delta,L)-D_y^2(\tau,L)$ and $\Delta M^2\equiv M^2(\tau,\delta,L)-M^2(\tau,L)$ obey
\begin{equation}
\label{eq:operator9}
\Delta D_y^2(\tau,\delta)\simeq L^{-d} \delta \tau^{\frac{d}{z}-\frac{2\beta}{\nu z}+\frac{1}{\tilde{\nu} \tilde{z}}},
\end{equation}
and
\begin{equation}
\label{eq:operator10}
\Delta M^2(\tau,\delta)\simeq L^{-d} \delta \tau^{\frac{d}{z}-\frac{2\beta}{\nu z}+\frac{1}{\tilde{\nu} \tilde{z}}},
\end{equation}
respectively. By analyzing the dependence of $\Delta D_x^2$, $\Delta D_y^2$ and $\Delta M^2$ on $\tau$, we can determine the specific choices of $\tilde{\nu}$ and $\tilde{z}$ for different quantities. For clarification, we enumerate the value of all possible combinations of $1/\tilde{\nu}\tilde{z}$: $1/\nu z\simeq 2.198$, $1/\nu' z=1/\nu z_u\simeq 1.572$ and $1/\nu' z_u\simeq 1.125$.
In the following calculations, we choose $\delta=\pm 0.005$. We have checked that for $\delta=\pm 0.01$, there is no qualitative difference in the universal properties.

At first, we study the case for $\delta<0$ with the ground state sitting in the VBS side. From Fig.~\ref{fig:initvbsgg} (a), one finds that for $\delta=-0.005$, $\Delta D^2_x$ evolves according to $\Delta D^2_x\propto \tau^{0.34}$. This exponent is close to $-2\beta/\nu z_u+1/\nu' z_u\simeq 0.246$. \revone{The deviation between the result in Fig.~\ref{fig:initvbsgg}(a) and the value $0.246$ can result from finite size effect.}
In addition, Figs.~\ref{fig:initvbsgg} (b) and (c) shows that $\Delta D_y^2\propto \tau^{1.93}$ and $\Delta M^2\propto \tau^{1.844}$. Both their exponents are close to $d/z-2\beta/\nu z+1/\nu' z_u\simeq 1.897$. These results show that for $D_x^2$, $D_y^2$ and $M^2$, their off-critical-point effects in the short-time stage are all controlled by $\delta\tau^{1/\nu' z_u}$. Since \revtwo{$\xi'\propto |\delta|^{-\nu'}$} and $\xi\propto \tau^{1/z_u}$, the term $\delta\tau^{1/\nu' z_u}$ \revone{combines the two arguments and reflects} an effect induced by the interplay of two length scales.

Then, let us turn to the case for $\delta>0$, in which the system is relaxed from the VBS initial state to a AFM ground state.
% The ground state is in the N\'{e}el phase in this case.
From Fig.~\ref{fig:initvbsgg} (d), one finds that for $\delta=0.005$, $\Delta D^2_x$ evolves according to $\Delta D^2_x\propto \tau^{0.32}$. This exponent is close to $-2\beta/\nu z_u+1/\nu' z_u \simeq 0.246$, similar to the case for $\delta<0$. In contrast, for $\Delta D_y^2$ and $\Delta M^2$, we find that $\Delta D_y^2\propto \tau^{2.23}$ and $\Delta M^2\propto \tau^{2.26}$, as shown in Figs.~\ref{fig:initvbsgg} (e) and (f). Both exponents are close to $d/z-2\beta/\nu z+1/\nu z_u$ or $d/z-2\beta/\nu z+1/\nu' z$, \revone{which are close to 2.344}. These results show that for $\delta>0$, the off-critical-point effects for $\Delta D_x^2$, $\Delta D_y^2$ and $\Delta M^2$ are controlled by different scaling variables: $\delta\tau^{1/\nu' z_u}$ dominates in $\Delta D_x^2$, while $\delta \tau^{1/\nu z_u}$ or $\delta \tau^{1/\nu' z}$ dominates in $\Delta D_y^2$ and $\Delta M^2$.

In addition, for $\Delta D_y^2$ and $\Delta M^2$, from the present numerical results, one cannot clarify which one of $\delta \tau^{1/\nu z_u}$ and $\delta \tau^{1/\nu' z}$ dominates the scaling form for $\delta>0$. Here we argue that both should be taken into account in the short-time stage. For the former, both $\delta^{-\nu}$ and $\tau^{1/z_u}$ are related to the usual length scale $\xi$, and $\delta \tau^{1/\nu z_u}$ represents the ratio between the contributions from $\delta$ and $\tau$ to $\xi$; while for the latter, both $\delta^{-\nu'}$ and $\tau^{1/z}$ are related to the confinement length $\xi'$, and $\delta \tau^{1/\nu' z}$ represents the ratio between the contributions from $\delta$ and $\tau$ to $\xi'$. These ratios can have similar order of magnitude in the universal short-time stage for finite $\delta$.
\revtwo{
  However, in the long-time stage, $\delta \tau^{1/\nu z_u}$ should dominate due to the asymmetric appearance of the dangerously irrelevant scaling variable. In the equilibrium situation, the dangerously irrelevant scaling variable only appear on the VBS side, causing scaling behaviors as a result of the interplay of the usual length scale $\xi$ and the additional confinement length scale $\xi'$.
  When relaxed to the AFM side, there is no dangerously irrelevant scaling variable in the equilibrium AFM ground state. Therefore, only the usual length scale, namely $\delta\tau^{1/\nu z_u}$, should governs the scaling form.
  }

  % However, in the long-time stage, $\delta \tau^{1/\nu z_u}$ should dominates since in the equilibrium situation there is no dangerously irrelevant scaling variable in the AFM side.

\revone{
Thus, we conclude the choice of $1/\tilde{\nu}\tilde{z}$ for different cases: 

(i) when $\delta<0$, for $\Delta D_x^{2}$, $\Delta D_y^{2}$ and $\Delta M^{2}$, $1/\tilde{\nu}\tilde{z}$ should be $1/\nu'z_u$, which reflects the interplay of two length scales. 

(ii) when $\delta>0$, for $\Delta D_x^{2}$, $1/\tilde{\nu}\tilde{z}$ is $1/\nu'z_u$ as well while for $\Delta D_y^{2}$ and $\Delta M^{2}$, $1/\tilde{\nu}\tilde{z}$ can be $1/\nu z_u$ or $1/\nu'z$. 
}

\subsection{Off-critical-point dynamic scaling from the AFM initial state}

\begin{figure}[tbp]
\centering
  \includegraphics[width=\linewidth,clip]{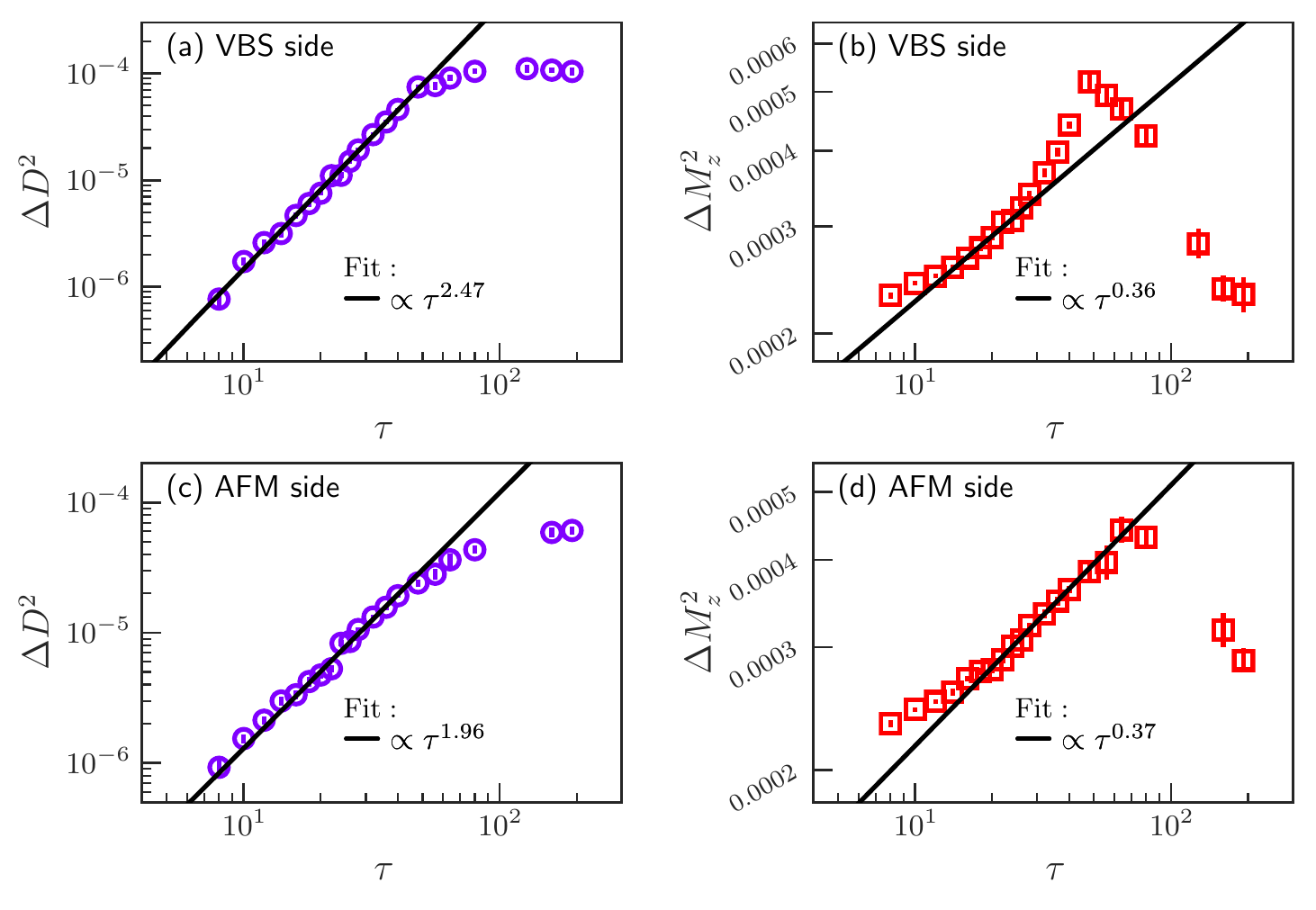}
  \vskip-3mm
  \caption{Relaxation dynamics away from the critical point with the AFM initial state. Top row: for $\delta>0$, curves of $\Delta D^2$ and $\Delta M_z^2$ versus $\tau$ are plotted in (a) and (b), respectively. Bottom row: for $\delta<0$, curves of $\Delta D^2$ and $\Delta M_z^2$ versus $\tau$ are plotted in (c) and (d), respectively. Power law fittings are implemented for all curves in the universal short-time stage. Double-logarithmic scales are used.
  }
  \label{initafgg}
\end{figure}

From the saturated AFM initial state, similar analyses give that for a small $\delta$, in the universal short-time stage, $\Delta D^2$ obeys the following scaling relation
\begin{equation}
\label{eq:operator11}
\Delta D^2(\tau,\delta)\simeq L^{-d} \delta \tau^{\frac{d}{z}-\frac{2\beta}{\nu z}+\frac{1}{\tilde{\nu} \tilde{z}}},
\end{equation}
and $\Delta M_z^2$ obeys
\begin{equation}
\label{eq:operator12}
\Delta M_z^2(\tau,\delta)\simeq \delta\tau^{-\frac{2\beta}{\nu z_u}+\frac{1}{\tilde{\nu} \tilde{z}}}.
\end{equation}
to the leading order of $\delta \tau^{1/\tilde{\nu} \tilde{z}}$.

Firstly, we study the case for $\delta<0$ with the ground state sitting in the VBS side. From Fig.~\ref{initafgg} (a), one finds that for $\delta=-0.005$, $\Delta M_z^2$ evolves according to $\Delta M_z^2\propto \tau^{0.36}$. This exponent is close to $-2\beta/\nu z_u+1/\nu' z_u\simeq 0.246$, similar to the scaling relation of $\Delta D^2_x$ from the VBS initial state. In contrast, from Fig.~\ref{initafgg} (b), we find that $\Delta D^2\propto \tau^{2.47}$ with its exponent close to $d/z-2\beta/\nu z+1/\nu z_u$ and $d/z-2\beta/\nu z+1/\nu' z$, \revone{both close to 2.344}. Similar to the case of $\Delta M^2$ with $\delta>0$ from the VBS initial state, we argue that both $\delta \tau^{1/\nu z_u}$ and $\delta \tau^{1/\nu' z}$ should be taken into account in the universal short-time stage.
\revtwo{
  % However, in the long-time stage, different from the case when the system is relaxed from the saturated VBS initial state to the AFM side, from the saturated AFM initial state to the VBS side, even with the presence of the dangerously irrelevant scaling variable, only one scale should governs. Here, the dominating scaling varialbe should be $\delta \tau^{1/\nu' z}$, since in equilibrium, $\xi'$ diverges faster than $\xi$ when $\delta$ is small as $\nu/\nu'<1$.
  However, in the long-time stage, different from the case when the system is relaxed from the saturated VBS initial state to the AFM side,
  from the saturated AFM initial state to the VBS side, besides $\delta \tau^{1/\nu z_u}$, $\delta\tau^{1/\nu' z}$ should also play impotant roles.
}
% However, differently, in this case, in the long-time stage, $\delta \tau^{1/\nu' z}$ should dominate since for the equilibrium situation in the VBS side $\xi'$ diverges faster than $\xi$ when $\delta$ is small.

Then, we turn to the case for $\delta>0$ and the ground state is in the N\'{e}el phase. From Fig.~\ref{initafgg} (c), one finds that for $\delta=0.005$, $\Delta M_z^2$ evolves according to $\Delta M_z^2\propto \tau^{0.37}$. This exponent is close to $-2\beta/\nu z_u+1/\nu' z_u\simeq 0.246$, similar to previous scaling relation of $\Delta M_z^2$ for $\delta<0$. Moreover, Fig.~\ref{initafgg} (d) shows that $\Delta D^2\propto \tau^{1.97}$ with the exponent close to $d/z-2\beta/\nu z+1/\nu' z_u\simeq 1.897$. These results show that for $\Delta D^2$ and $\Delta M_z^2$, the off-critical-point effects in the universal short-time stage is controlled by $\delta\tau^{1/\nu' z_u}$ when $\delta<0$.

\revone{
Thus, we conclude the choice of $1/\tilde{\nu}\tilde{z}$ for different cases: 

(i) when $\delta<0$, for $\Delta M_z^{2}$, $1/\tilde{\nu}\tilde{z}$ is $1/\nu'z_u$ while for $\Delta D^2$, the choice can be $1/\nu z_u$ or $1/\nu'z$. 

(ii) when $\delta>0$, for $\Delta M_z^{2}$ and $\Delta D^{2}$,  $1/\tilde{\nu}\tilde{z}$ should be $1/\nu'z_u$.

% Comparing with the case of VBS initial state , one finds that, when relaxing to the same $\delta$ from different initial ordered state , the difference of dominant order parameter no longer share the same form. So do the complementary ones. The dual dynamic scaling breaks down on moving away from the critical point.

}

\subsection{Off-critical-point dynamic scaling from the disordered initial state}

\begin{figure}[tbp]
\centering
  \includegraphics[width=\linewidth,clip]{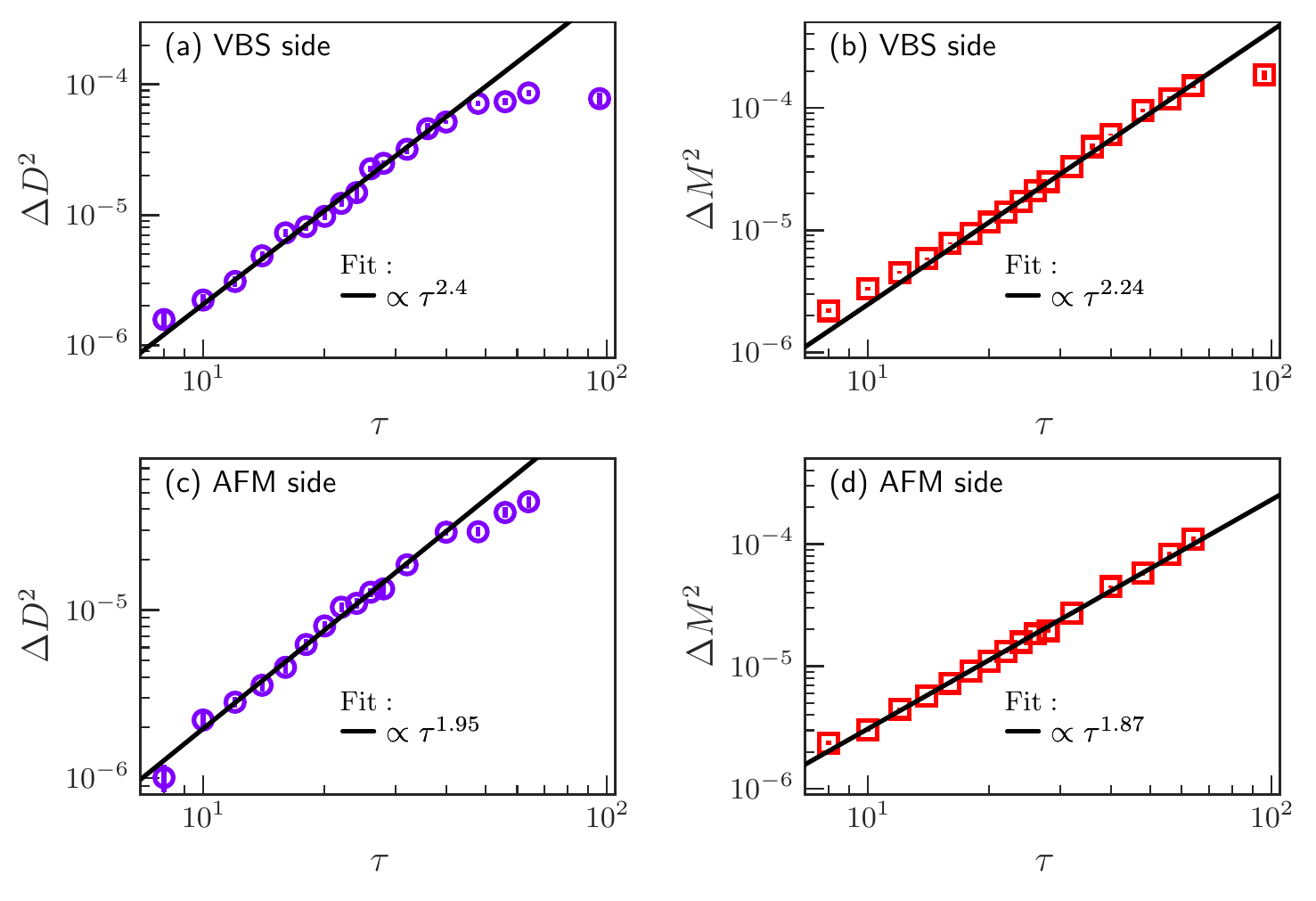}
  \vskip-3mm
  \caption{Relaxation dynamics away from the critical point with the disordered initial state. Top row: for $\delta>0$, curves of $\Delta D^2$ and $\Delta M^2$ versus $\tau$ are plotted in (a) and (b), respectively. Bottom row: for $\delta<0$, curves of $\Delta D^2$ and $\Delta M^2$ versus $\tau$ are plotted in (c) and (d), respectively. Power law fittings are implemented for all curves in the universal short-time stage. Double-logarithmic scales are used.
  }
  \label{initpmgg}
\end{figure}

For the disordered initial state, similar analyses give that for a small $\delta$, in the universal short-time stage, $\Delta D^2$ obeys
\begin{equation}
\label{eq:operator13}
\Delta D^2(\tau,\delta)\simeq L^{-d} \delta \tau^{\frac{d}{z}-\frac{2\beta}{\nu z}+\frac{1}{\tilde{\nu} \tilde{z}}},
\end{equation}
and $\Delta M^2$ obeys
\begin{equation}
\label{eq:operator14}
\Delta M^2(\tau,\delta)\simeq L^{-d} \delta \tau^{\frac{d}{z}-\frac{2\beta}{\nu z}+\frac{1}{\tilde{\nu} \tilde{z}}},
\end{equation}
to the leading order of $\delta \tau^{1/\tilde{\nu} \tilde{z}}$.

For $\delta<0$ with the VBS ground state, Figs.~\ref{initpmgg} (a)-(b) shows that for $\delta=-0.005$, $\Delta D^2$ and $\Delta M^2$ evolve according to $\Delta D^2\propto \tau^{2.4}$ and $\Delta M^2\propto \tau^{2.24}$, respectively. The two exponents are close to $d/z-2\beta/\nu z+1/\nu z_u$ and $d/z-2\beta/\nu z+1/\nu' z$, both close to $2.344$. As discussed above, both $\delta \tau^{1/\nu z_u}$ and $\delta \tau^{1/\nu' z}$ should make contributions in the short-time regime. For $\delta>0$ with the AFM ground state, from Fig.~\ref{initpmgg} (c)-(d), one finds that $\Delta D^2\propto \tau^{1.95}$ and $\Delta M^2\propto \tau^{1.87}$ when $\delta=0.005$. Both these exponents are close to $d/z-2\beta/\nu z+1/\nu' z_u=1.897$, demonstrating the off-critical-point effects are mainly contributed by $\delta \tau^{1/\nu' z_u}$.

\revone{
  Thus, we conclude that with the disordered initial state,

  (i) when $\delta<0$, for both $\Delta M^{2}$ and $\Delta D^{2}$, $1/\tilde{\nu}\tilde{z}$ can be $1/\nu z_u$ or $1/\nu'z$. 

  (ii) when $\delta>0$, for both $\Delta M^{2}$ and $\Delta D^{2}$, $1/\tilde{\nu}\tilde{z}$ is $1/\nu'z_u$.
  }

\subsection{Breakdown and vestige of the dual dynamic scaling}

From the above results, we find that the dual dynamic scaling breaks down under the rotation between the superspin components when the tuning parameter is away from its critical point. In the long-time limit, the system reaches its equilibrium ground state, in which the asymmetric scaling properties between the N\'{e}el order and the VBS order appear~\cite{Delamotte2015prl,Louj2007prl} and meanwhile the emergent symmetry fades out~\cite{Sandvik2009prb1}. In the short-time stage, from the ordered initial state, for a fixed $\delta\neq 0$, under the rotation between $D$ and $M$, one finds that the dual dynamic scaling also breaks down. Although \revone{the dominant} order parameters of the ordered initial states obey similar scaling behavior, \revone{the complementary} order parameters behave quite differently.
When relaxed from the saturated VBS initial state to the VBS side, i.e. $\delta<0$,\revone{for the complementary order parameter $M$}, its off-critical-point effects $\Delta M^2$ is dominated by $\delta \tau^{d/z-2\beta/\nu z+1/\nu' z_u}$; as the dual case, from the AFM initial state, in contrast, \revone{for the complementary order parameter $D$}, its off-critical-point effects $\Delta D^2$ is dominated by $\delta \tau^{d/z-2\beta/\nu z+1/\nu' z}$ and\revone{/or} $\delta \tau^{d/z-2\beta/\nu z+1/\nu z_u}$.
Similarly, when relaxed from the saturated VBS initial state to the AFM side, i.e. $\delta>0$, $\Delta M^2$ is dominated by $\delta \tau^{d/z-2\beta/\nu z+1/\nu' z}$ and\revone{/or} $\delta \tau^{d/z-2\beta/\nu z+1/\nu z_u}$; as its dual case, from the AFM initial state, in contrast, $\Delta D^2$ is dominated by $\delta \tau^{d/z-2\beta/\nu z+1/\nu' z}$. This demonstrates that the dual dynamic scaling breaks when the parameter is tuned away from the critical point.

However, strikingly, for the ordered initial states, we find that the dual dynamic scaling in the universal short-time stage can restore once the transformation $M\Leftrightarrow D$ and $\delta\Leftrightarrow -\delta$ are simultaneously implemented. Under this joint transformation, the relaxation dynamics of the VBS (N\'{e}el) order parameter for $\delta<0$ with the VBS intial state can be dualized to the relaxation dynamics of the N\'{e}el (VBS) order parameter for $\delta>0$ with the N\'{e}el initial state, and vice versa. Accordingly, the scaling behaviors of both the \revone{dominant and the complementary} order parameters keep invariant under this joint dual transformation. \revone{In this sense, the dual dynamic scaling is maintained.}

The vestige of the dual dynamic scaling in the short-time stage demonstrates that the nonequilibrium dynamics can even exhibit much higher symmetry than the equilibrium state. \revone{Therefore, unlike the fade of the emergent symmetry on running away from the critical point in equilibrium, the dual dynamic scaling can survive up to some finite $\delta$ under the joint dual transformation.} A possible reason is that the nonequilibrium dynamics mixes various elements together to form a state which is covariant under the dual transformation, although each single element is asymmetric under the same condition. For instance, relaxing from the saturated VBS initial state to the VBS side with a finite $\delta<0$, the relaxation dynamics mixes the the VBS background, the VBS ground state, and the topological defects with spinons living at the vortices of the VBS domain walls; while relaxing from the AFM initial state to the AFM side with a finite $\delta>0$, the relaxation dynamics mixes the local fluctuations with the background in the AFM pattern, the AFM ground state, and the topological defects with quadrupled monopole excitations. The interplay between these elements may give rise to the dual dynamic scaling. However, from the disordered initial states, in the short-time stage, only global fluctuations contribute to $M^2$ and $D^2$. Thus the dual dynamic scaling in the short-time stage still exists for the transformation of $M\Leftrightarrow D$. However, it breaks down when $\delta\Leftrightarrow -\delta$ is added simultaneously, since in this case, the ground state information, which breaks the emergent symmetry of the superspin rotation, should be simultaneously included. 

\section{\label{experimentdis}Possible experimental realization}

\begin{figure}[htbp]
\centering
  \includegraphics[width=\linewidth,clip]{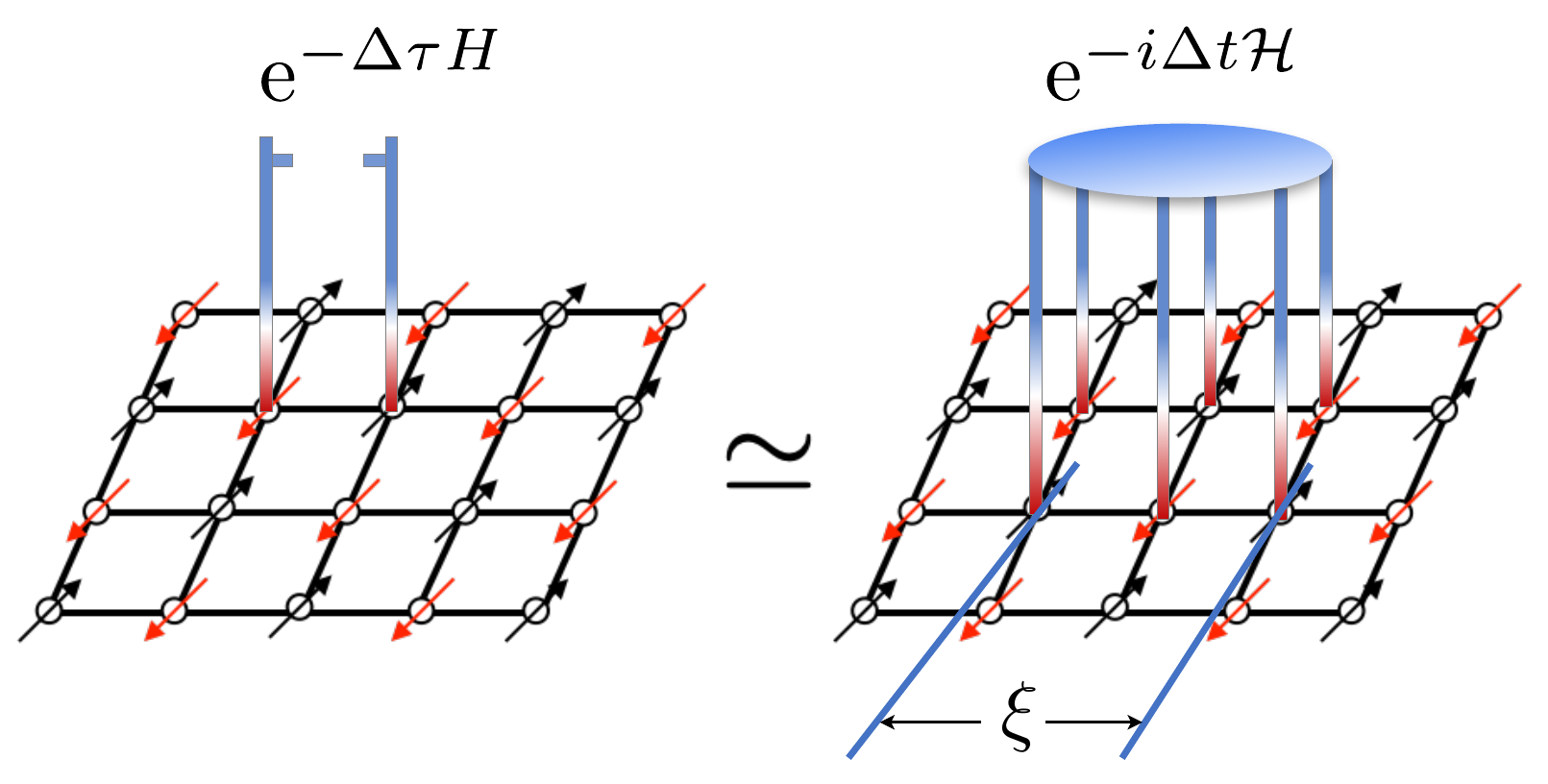}
  \vskip-3mm
  \caption{Realization of imaginary-time relaxation dynamics via unitary quantum gates.
  }
  \label{reims}
\end{figure}

Recently, it was shown that quantum computers has become a new vivifying platform to realizing various `experiments' ranging from high-energy physics~\cite{Martinez2016nature,Bauer2021prl} to condensed matter physics~\cite{Satzinger2021science,Semeghini2021science}. In particular, nonequilibrium quantum critical dynamics has been observed in the D-wave devices~\cite{Weinberg2020prl} and the noisy intermediate-scale quantum device based on the Rigetti superconducting quantum chip~\cite{JMoore2021arxiv}. Moreover, imaginary-time relaxation was also implemented in the various devices as a promising approach to find the ground state~\cite{Motta2020naturephyscis}. As illustrated in Fig.~\ref{reims}, imaginary-time evolution in a small time interval $\Delta \tau$ with the Hamiltonian $H$ can be approximated by the real-time unitary evolution with the auxiliary Hamiltonian $\mathcal{H}$. By minimizing the following residual form,
\begin{equation}
\label{eq:operator15}
{\rm Res}\equiv\left\| \frac{e^{-\Delta \tau H}|\psi\rangle}{\langle\psi|e^{-2 \Delta \tau H}|\psi\rangle}-e^{-i\Delta t\mathcal{H}}|\psi\rangle\right\|^2,
\end{equation}
one can determine $\mathcal{H}$~\cite{Motta2020naturephyscis,Nishi2021njp}.
\revtwo{
  It was shown that the interaction length in $\mathcal{H}$ is only required to have the same order of magnitude as the correlation length $\xi$.
}
For our present case, the initial state is set as the uncorrelated state with vanishing correlation length~\cite{Motta2020naturephyscis}. Thus it is promising that our results can be detected in these systems in the future.

\section{\label{summaryd}Summary}
In summary, we have studied the nonequilibrium imaginary-time dynamics of the DQCP in the $2$D $J$-$Q_2$ model. At the critical point, we have generalized the emergent symmetry in equilibrium to the dual dynamic scaling in the relaxation process. In particular, we have found that in the relaxation dynamics from the ordered initial states, although the dominant and the complementary order parameters are controlled by different length scales, they can be \revtwo{dualized} to each other under the rotation transformation in the superspin space. In addition, we have shown that for the disordered initial state the dual dynamic scaling dictates that both the N\'{e}el and the VBS order parameter obey the same scaling form. By comparing the results of the $J$-$Q_3$ model~\cite{Shu2021prl}, we have confirmed that the dual dynamic scaling is a universal nonequilibrium phenomenon in the DQCP that separates the N\'{e}el order and the VBS order. Furthermore, we have studied the relaxation dynamics away from the critical point and found that the dual dynamic scaling can even exist when the emergent symmetry in the ground state has been broken. We have attributed the appearance of the dual dynamic scaling at and near the DQCP to the superposition of information included in the initial state, ground state and the low-energy excited states. In addition, we have discussed the possible experimental realizations in the fast-developing programmable quantum devices.

In this paper, we have only focused on the situation where the initial state is at its fixed points. In usual critical point with one single length scale, it was shown that the initial state can induce interesting scaling behaviors~\cite{Janssen1989,Zhengb1996prl,Yins2014prb,Yins2014pre,Shu2017prb,Shu2020prb}. In our previous work~\cite{Shu2021prl}, we have shown that in DQCP, the critical initial slip exponent is negative, while in usual phase transitions, this exponent is positive. Thus it is instructive to explore more general effects induced by different initial states.

Moreover, at present, directly attacks on the nonequilibrium dynamics in two-dimensional quantum systems is still confronted with sever difficulties. Although the direct analytic continuations seem inadequate to obtain the results of the real-time dynamics, it has been shown that some universal behaviors are shared in both the imaginary-time and real-time dynamics~\cite{Polkovnikov2011prb,De_Grandi_2013,PolkovnikovSandvik2013prb,KOLODRUBETZ20171}. In addition, it was shown that the real-time relaxation has the similar scaling relations with the imaginary-time relaxation, but with different critical exponent~\cite{Polkovnikov2013rephprl,Gagel2014prl,Maraga2015pre,Mitra2015prb,Mitra2016prb,Marino2017prl,Yin2019prl,YinJian2021prb,Jad2021pre}. Therefore, it is expected that our present work should provide insights to the real-time dynamics of the DQCP.

\section*{Acknowledgements}
Y.R.S. acknowledges support from the National Natural Science Foundation of China (Grants No. 11947035 and No. 12104109), the Science and Technology Projects in Guangzhou (202102020933). S.Y. is supported by the National Natural Science Foundation of China (Grant No. 12075324).

\bibliography{ref}

\end{document}